\documentclass{aa} 

\usepackage{graphicx}
\usepackage{txfonts}
\usepackage[]{natbib}
\usepackage{rotating}
\usepackage{pdflscape}

        \newcommand{\msun}{\ensuremath{\mathrm{M}_{\odot}}}
        \newcommand{\lsun}{L$_{\odot}$}
        \newcommand{\sfr}{M$_{\odot}$ yr$^{-1}$}
        \newcommand{\lagn}{${\rm L_{AGN} }$}
        
    \newcommand{\oiii}{[O{\sc iii}]$\lambda$5007}
    \newcommand{\aco}{$\alpha_{\rm CO}$}
    \newcommand{\mgas}{${\rm M_{\rm gas}}$}
    \newcommand{\mbh}{${\rm M_{\rm BH}}$}
        \newcommand{\mstar}{${\rm M_{\star}}$}
        \newcommand{\ha}{${\rm H\alpha}$}
        \newcommand{\hb}{${\rm H\beta}$}
        \newcommand{\ujy}{$\mu$Jy}
        \newcommand{\tscale}{$\tau_{\rm dep}$}
        \newcommand{\qz}{2QZJ0028}      
        \newcommand{\lb}{LBQS0109}
        \newcommand{\hbs}{HB8903}
        \newcommand{\rco}{r$_{31}$}

\begin{document}



\title{AGN feedback on molecular gas reservoirs in quasars at $z\sim2.4$}


\author{S.~Carniani\inst{1,2},
          A.~Marconi\inst{3,4},
          R.~Maiolino\inst{1,2},
          C.~Feruglio\inst{5},
          M.~Brusa\inst{6,7},
          G.~Cresci\inst{4},
          M.~Cano-D\'iaz\inst{8},
          C.~Cicone\inst{9},
          B.~Balmaverde\inst{10},
          F.~Fiore\inst{11},
          A.~Ferrara\inst{10},
                  S.~Gallerani\inst{10},
          F.~La~Franca\inst{12},
          V.~Mainieri\inst{13},
          F.~Mannucci\inst{4},
          H.~Netzer\inst{14},
          E.~Piconcelli\inst{11},
          E.~Sani\inst{16},     
          R.~Schneider\inst{10},
              O.~Shemmer\inst{15},
              L.~Testi\inst{13,4,17}
              }    
\institute{Cavendish Laboratory, University of Cambridge, 19 J. J. Thomson Ave., Cambridge CB3 0HE, UK 
\and
Kavli Institute for Cosmology, University of Cambridge, Madingley Road, Cambridge CB3 0HA, UK 
\and
Dipartimento di Fisica e Astronomia, Universit\`a di Firenze, Via G. Sansone 1, I-50019, Sesto Fiorentino (Firenze), Italy 
\and
INAF - Osservatorio Astrofisico di Arcetri, Largo E. Fermi 5, I-50125, Firenze, Italy 
\and
INAF - Osservatorio Astronomico di Trieste, via G. Tiepolo 11, I34124 Trieste, Italy
\and
Dipartimento di Fisica e Astronomia, Universit\`a di Bologna, viale Berti Pichat 6/2, 40127 Bologna, Italy  
\and
INAF - Osservatorio Astronomico di Bologna, via Ranzani 1, 40127 Bologna, Italy 
\and
Instituto de Astronom\'{\i}a, Universidad Nacional Aut\'onoma de M\'exico, Apartado Postal 70-264, Mexico D.F., 04510 Mexico 
\and    
INAF-Osservatorio Astronomico di Brera, via Brera 28, I-20121, Milano, Italy
\and
SNS - Scuola Normale Superiore, Piazza dei Cavalieri 7, I-56126 Pisa, Italy 
\and
INAF - Osservatorio Astronomico di Roma, via Frascati 33, 00040 Monteporzio Catone, Italy 
\and 
Dipartimento di Matematica e Fisica, Universit\`a Roma Tre, via della Vasca Navale 84, I-00146 Roma, Italy 
\and
European Southern Observatory, Karl-Schwarzschild-str. 2, 85748 Garching bei M\"unchen, Germany 
\and
School of Physics and Astronomy, The Sackler Faculty of Exact Sciences, Tel-Aviv University, Tel-Aviv 69978, Israel 
\and
Department of Physics, University of North Texas, Denton, TX 76203, USA
\and
European Southern Observatory, Alonso de Cordova 3107, Vitacura, Santiago, Chile 
\and
Excellence Cluster 'Universe', Boltzmannstr. 2, D-85748 Garching bei M\"unchen, Germany 
}

\abstract{
We present new ALMA observations aimed at mapping molecular gas reservoirs through the CO(3-2) transition in three quasars at $z\simeq2.4$, LBQS 0109+0213, 2QZ J002830.4-281706, and [HB89] 0329-385.
Previous \oiii\ observations of these quasars showed evidence for ionised outflows quenching star formation in their host  galaxies.
Systemic CO(3-2) emission has been detected only in  one quasar, LBQS 0109+0213,
where  the CO(3-2) emission is spatially anti-correlated with the ionised outflow,  suggesting that  most of the molecular gas may have been dispersed or heated in the region swept by the outflow. 
In all three sources, including the one detected in CO, our constraints on the molecular gas mass indicate a significantly reduced reservoir compared to main-sequence galaxies at the same redshift, supporting a  negative feedback scenario.
%
In the quasar 2QZ J002830.4-281706, we tentatively detect an emission line blob blue-shifted by $v\sim-2000$~km/s with respect to the galaxy systemic velocity and spatially offset by 0.2\arcsec\ (1.7 kpc) with respect to the ALMA continuum peak.  Interestingly, such emission feature is coincident in both velocity and space with the ionised outflow as seen in \oiii. 
This tentative detection must be confirmed with deeper observations but, if real, it could represent the molecular counterpart of the ionised gas outflow driven by the Active Galactic Nucleus (AGN). 
Finally, in all ALMA maps we detect the presence of serendipitous line emitters within a projected distance $\sim 160$ kpc from the quasars.
By identifying these features with the CO(3-2) transition, we find that the serendipitous line emitters would be located within |$\Delta v$|$<$500 km/s from the quasars, hence suggesting an overdensity of galaxies in two out of three quasars.
}


\authorrunning{Carniani et al.}
\titlerunning{AGN feedback on molecular gas reservoirs in quasars at $z\sim2.3$}
\maketitle



\section{Introduction}

Both the growth of super-massive black holes (SMBHs) and star formation history of galaxies are regulated by the supply of cold gas  available in the host. The molecular gas reservoir can be replenished  through either accretion of cold gas from the halo or  wet mergers.

Intense bursts of star formation, such as those observed in sub-millimetre galaxies (SMGs), and  in the host galaxies of powerful active galactic nuclei (AGN) can be induced by mergers, interactions and disk instabilities \cite[see, e.g.][]{Alexander:2012}. 
 Several studies have suggested that SMGs and quasars (QSOs) represent two distinct stages of galaxy evolution \citep[e.g.][]{Hopkins:2008}. SMGs  would correspond to the starburst phase when galaxies are dust obscured and therefore emit mainly  at far-infrared wavelengths. QSOs are unobscured systems  where the gas has been expelled by energetic outflows, which eventually quench both SMBH growth and star formation (SF) \citep{Di-Matteo:2005}.

The discovery of  ubiquitous massive, powerful galaxy-wide outflows  in QSO host galaxies supports the QSO feedback scenario  depicted above.  Studies based on millimetre observations of local QSO hosts have estimated molecular outflow mass-loss rates exceeding the star formation rates by almost two orders of magnitude in the most powerful sources \citep{Cicone:2014}.  Massive outflows can deplete the host galaxies of their  cold gas content in approximately a few Myrs, that is, on timescales even shorter than the depletion time scales due to gas consumption by star formation \citep{Maiolino:2012, Cicone:2014}. However, although  these observations are in overall agreement with AGN feedback models \citep[see][and references therein]{Fabian:2012}, we are still missing the smoking gun evidence that the  AGN-driven outflows are effectively quenching star formation: what we are seeking is a clear and unambiguous indication that star formation is indeed inhibited in the galaxy regions swept by the outflows.

Several SINFONI/VLT observations of $z\sim1.5-2.5$ QSOs indicate the presence of fast, galaxy-wide ionised outflows with a conical morphology that are spatially anti-correlated with  the brightest actively star forming region in the host galaxy \citep{Cano-Diaz:2012, Carniani:2015a, Cresci:2015, Carniani:2016}.
These results suggest that the fast winds are  simultaneously expelling gas from the host galaxies and quenching star formation in the region  swept by the outflow. However,  we note that, at optical wavelengths, observations may be affected by differential extinction effects,  and so we cannot fully rule out the presence of obscured emission powered by star formation in the region affected by the ionised outflow.  In conclusion, it is still debated whether the  observed absence of  star formation signatures in the outflow region is  real and related to gas depletion  by feedback or if  it is caused by dust obscuration.

In this context, observations at (sub-)millimetre wavelengths are crucial to definitely establish whether star formation is inhibited by fast outflows.  Through the carbon monoxide (CO) line emission  we can directly trace the cold molecular gas that fuels the star formation activity, and thus confirm or reject negative-feedback scenarios.
 \cite{Brusa:2015a} reported the detection of CO(3-2) emission with the IRAM Plateau de Bure Interferometer (PdBI) in XID2028, one of the $z\sim2$ QSOs exhibiting spatial anti-correlation between narrow \ha\ emission, tracing  star formation, and ionised AGN-driven outflows \citep{Cresci:2015}. The  modest molecular mass  inferred from the CO(3-2) line detection indicates that the gas in the host galaxy has been already depleted  or dispersed by QSO feedback.
However the angular resolution ($\sim4\arcsec$) of these PdBI observations is not sufficient to spatially resolve the CO(3-2) emission in  XID2028,  hence not allowing an accurate determination of the location of the molecular gas reservoir with respect to the ionised outflow \citep{Brusa:2015a}.

We have  recently undertaken an Atacama Large Millimetre/submillimetre Array (ALMA) programme targeting two QSOs of the sample by \cite{Carniani:2015a}, LBQS0109+0213 (hereafter \lb),  2QZJ002830.4-281706 (hereafter \qz),  and [HB89] 0329-385 (hereafter HB8903)  in which the  spatial distribution of their narrow \ha\ and \oiii\ emissions  with respect to the location of the \oiii\ outflow supports a negative-feedback scenario \citep{Cano-Diaz:2012, Carniani:2016}. 
The aim of the ALMA observations presented in this paper is to map the molecular gas in the host galaxies  through the CO(3-2) emission (rest frequency $\nu_{\rm rest}=345.8$ GHz) and compare the spatial distribution of the molecular gas with that of the fast-outflowing ionised gas.
The paper is organised as follows: Sect.~\ref{sec:sample} describes the  target properties, and Sect.~\ref{sec:observations}  summarises the ALMA observations. In Sect.~\ref{sec:lbqs}, \ref{sec:2qzj0028}, and \ref{sec:hb8903} we present  our results on \lb, \qz, and HB8903, respectively. A discussion of the molecular gas content in  all host galaxies is included in Sect.~\ref{sec:lack} and, our conclusions are summarised in Sect.~\ref{sec:overdensity} and \ref{sec:summary}. 

In this work we adopt a $\Lambda$CDM cosmological model with ${\rm H_0 = 67.3 \  km \ s^{-1} \  Mpc^{-1} , \ \Omega_M = 0.315,  \ \Omega_{\Lambda} = 0.685}$ \citep{Planck-Collaboration:2014aa}. According to  this model, 1\arcsec\ at $z = 2.4$ corresponds to a  physical scale of 8.35 kpc.

\section{Sample selection }
\label{sec:sample}

The selected targets, \lb, \qz, and HB8903, are part of a large high-luminosity (L$>10^{47}$ erg/s) QSO sample at $z>2$  \citep{Shemmer:2004, Netzer:2004},  characterised by \oiii\ equivalent widths of $EW>10$~\AA\ and  relatively bright in the H band, that is, H$<16.5$~mag. The properties of the three targets are listed in Table~\ref{tab:summary}. 

\begin{table*}
\caption{Properties of the three quasars from literature}           
\label{tab:summary}
\begin{tabular}{l c c c c c  c c  c c c}    
                         \\
\hline
 & & & &  & &  & \multicolumn{4}{c}{Ionised outflow properties}  \\
QSO &   $z$ & RA & DEC  & Log$_{10}$($\frac{L_{\rm AGN}}{\rm erg/s})$ & $M_{\rm BH}$ & SFR & $v$   & $R$  & $M_{\rm [OIII]}$      & $  \dot M_{\rm [OIII]}$  \\

   &  &  & & & {\small[$10^{10}$ \msun]} & {\small[\sfr]} & {\small[km/s]} & {\small[kpc]} & {\small[$10^7$ \msun]} & {\small[\sfr ]}  \\
 {\small(1)}  & {\small(2)} & {\small(3)} & {\small (4)} & {\small(5)} & {\small(6)} & {\small(7)} & {\small(8)} & {\small(9)} & {\small(10)} & {\small(11)}\\

\hline
LBQS0109  & 2.35  & 01:12:16.99 & +02:29:47.7   &  47.5 & 1.0 & 50 & 1850 & 0.4 & 1.2 & 60    \\

2QZJ0028 & 2.40  &  00:28:30.42 &  -28:17:05.4    & 47.3 & 1.2 & 100 & 2300 & 0.7 & 3.8 & 140 \\

HB8903 & 2.44  &  03:31:06.41 &  -38:24:04.6    & 47.5 & 1.3 & 90 & 1450 & 1.9 & 0.7 & 6 \\
\hline
                
\end{tabular}   
\tablefoot{ (1) ID of the object (2) Redshifts estimated from  the narrow \oiii\  emission \citep{Carniani:2015a}. (3,4) Coordinates (J2000.0). (5) AGN bolometric luminosities derived by using the relation  \lagn \ $\sim 6\,\lambda L(\lambda5100\AA)$  from \cite{Marconi:2004}. (6) Black hole masses from \cite{Shemmer:2004} and \cite{Williams:2016}. (7) SFR estimated from  \ha\ emission \citep{Carniani:2016}. (8, 9) \oiii\ Outflow velocity and  inferred by using the spectro-astrometric method described in \cite{Carniani:2015a}. (10) Outflow masses inferred from \oiii\ by assuming  ${\rm T_e \sim 10^4}$ K and  ${\rm n_e \sim 500 cm^{-3}}$. (11) Outflow mass-loss rates calculated as $\dot{M}_{\rm o} = M_{\rm o}/\tau_{\rm dyn} =  M_{\rm o}v_{\rm o}/R_{\rm o}$
. }

\end{table*}

To investigate  the properties of the AGN-driven outflows,  we have observed the three QSOs in the H (1.45-1.85 $\mu$m) and K (1.95-2.45 $\mu$m) bands with the Spectrograph for INtegral Field Observations in the Near Infrared (SINFONI)  in seeing limited mode  (angular resolution $\sim0.6$\arcsec).
The kinematical analysis of the \oiii\ emission line revealed fast ($>1000$ km/s) ionised outflows extended a few kpc from the galaxy centre \citep{Cano-Diaz:2012, Carniani:2015a}.
In addition,  the nuclear spectrum of \qz, which has also been recently observed with SINFONI assisted by adaptive optics (angular resolution $\sim0.15\arcsec$), is  characterised  by a broad, blueshifted \hb\ absorption trancing nuclear outflowing gas with density higher than $10^9$ cm$^{-3}$ and velocity up to 10,000 km/s \citep{Williams:2016}.

Intriguingly, the presence of  extended outflows appears to be spatially anti-correlated with the narrow \ha\ emission component  tracing star formation in the host galaxy \citep{Cano-Diaz:2012, Carniani:2016}.
These results have been interpreted as evidence  for negative feedback in action,  where star formation is quenched in the region where AGN-driven outflows interact with the host galaxy.
 If excluding dust-extinction effects, the reduction of star formation activity in the outflow region can be caused by a lack of a substantial molecular gas reservoir, that may have been expelled by the outflow itself, or to heating and turbulence effects related to the feedback process that may lower the star formation efficiency of the gas.

\section{Observations}
\label{sec:observations}

LBQS0109, 2QZJ0028, and HB8903 were observed  at the Band~3 frequencies ($\sim100$ GHz, corresponding to $\lambda\sim3$~mm) with the ALMA array between July 2015 and July 2016. The  on-source time  was about 40~minutes with  40-44 12-m antennas for all sources. 
The antennas were distributed in a 
semi-compact configuration with a maximum baseline length of $\sim 1.5$ km. The average precipitable water vapour (PWV) values during the observations were 3~mm, 2.2~mm and 0.8~mmm for the three targets, respectively. 

The millimetre observations, carried out in frequency division mode, have a total bandwidth of 7.5 GHz divided into four spectral windows of $\sim1.875$ GHz with a channel width of 1.9 MHz ($\sim$ 5.7 km/s). 
One of the four spectral windows was tuned to the expected central frequency of the CO(3-2) line, that is, 103.2 GHz for LBQS0109,  101.6 GHz for 2QZJ0028, and  100.5 GHz for HB8903. The redshifts of the three QSOs were estimated from the  narrow \oiii\ and \ha\ emission lines \citep{Cano-Diaz:2012, Carniani:2016}.

The data were calibrated  using the CASA software version v4.5.2 \citep{McMullin:2007}. The phase calibrators were J0038-2459 and J0108+0135 for LBQS0109 and \qz, respectively. Ceres and J0238+166 were used as flux calibrators, while  bandpass calibrations were carried out through  the observations of J2258-2758 and J0238+1636. The flux, bandpass and phase calibrator for HB8903 was J0334-4008.
All final images were reconstructed by using the CASA task {\sc clean}. 

Continuum images at 3 mm were obtained using the line-free channels of the four spectral windows. 
By using a natural weighting we achieved  for all sources a sensitivity of 12 \ujy/beam for the first two QSOs and 18 \ujy/beam  for HB8903. The final images have an angular resolution of about 0.6\arcsec, which corresponds to 5 kpc at $z\sim2.4$, and a spatial scale per pixel of  0.1\arcsec. 

We subtracted the  continuum emission by fitting an UV~-~plane model to the line-free channels of each spectral window using the {\sc uvcontsub} task. We  generated the final cubes from the continuum subtracted data using the {\sc clean} task with the parameter {\sc weighting = briggs} and {\sc robust = 0.5}, which offers a compromise between high-resolution and highest sensitivity per beam.
In all sources, we achieved a 1$\sigma$ sensitivity of 240-280 \ujy/beam per spectral bin of 30 km/s with a beam size of about 0.6\arcsec$\times$0.5\arcsec. 
The angular resolution of ALMA images  matches well that of  the SINFONI observations ($\sim$0.6\arcsec). 

The source size and the flux density of the continuum emission of the three sources are inferred by fitting a 2D elliptical Gaussian profile to the visibility data in CASA by  using the {\sc uvmodelfit} task. The line properties were estimated in the image plane instead. 

In this work we are interested in comparing ALMA and SINFONI observations,  hence we verified the astrometry accuracy of both  datasets. The absolute positions of the  QSOs in the ALMA field are consistent with the Sloan-Digital-Sky-Survey (SDSS)  and Two-Micron-All-Sky-Survey (2MASS) positions within the astrometric uncertainty of about 0.1\arcsec.     
As no astrometric calibrations of SINFONI were observed for the three targets, we had to align the peak of the H- and K-band SINFONI continuum emission with the centroid position obtained from 2MASS images in the same bands (H and K).

\begin{table*}
\caption{Millimetre properties of \lb, \qz\ and HB8903.}           
\label{tab:alma}      
\centering          
\begin{tabular}{l c c c  }    
\\
\hline
 &  LBQS0109 &  \qz & HB8903  \\
\hline
\\
$\sigma_{\rm 3mm}$ [\ujy/beam] & 12 & 12 & 18 \\
S$_{\rm 3mm}$ [\ujy] & $160\pm16$  &  $168\pm14$  & $5692\pm12$\\

 Major-axis$_{\rm 3mm}$ [\arcsec]$^{a}$  & $0.5\pm0.1$ & - & $0.102\pm0.006$\\
 Axis-ratio$_{\rm 3mm}$$^{a}$  & $0.5\pm0.4$ & - & $ 1.00\pm0.06$ \\
 P.A.$_{\rm 3mm}$[\degr]$^{a}$  & $85\pm11$ & - & $-90\pm60$\\

M$_{\rm dust}$ [10$^{9}$ \msun]$^{b}$ & 0.5-0.8 & 0.6-0.9  & 20-30\\

$\lambda_{\rm CO(3-2)}$ [mm] & 2.9094$\pm$0.0004 & - & -  \\

$z_{\rm CO(3-2)}$  & 2.3558$\pm$0.0005 & -  & - \\

FWHM$_{\rm CO(3-2)}$ [km/s] & 400$\pm$60 & -  & - \\

S$_{\rm CO(3-2)} \Delta v$ [Jy km/s]$^{c}$ & $0.34\pm0.03$ & $<0.09$ & $<0.08$   \\

L$^\prime_{\rm CO(3-2)}$ [$10^{10}$ K km/s pc$^2$]$^{c}$ & $1.04\pm0.33$ & $<0.3$ & $<0.3$  \\

L$_{\rm CO(3-2)}$ [$10^{7}$ \lsun]$^{c}$ & $1.4\pm0.2$  &  $<0.4$ &  $<0.3$  \\


M$_{\rm gas}$(\aco=0.8) [$10^{10}$ \msun]$^{c,d}$ &  $0.8\pm0.5$ & $<0.2$ & $<0.2$ \\

M$_{\rm gas}$(\aco=4) [$10^{10}$ \msun]$^{c,d}$   &  $4.0\pm2.4$ & $<1.2$ & $<1.0$\\
\\
\hline
\\              
\end{tabular}   
\tablefoot{$^{a}$ Beam-deconvolved size estimated in the UV-plane by using the {\sc UVMODELFIT} task.$^{b}$ Under the assumption that the continuum emission at 3~mm is completely associated  to thermal dust continuum emission.
We assume a ${T_{d}}$=40-60 K and a $\beta=2.0$. $^{c}$ For \qz\ and HB8903, we assume a line width of 400 km/s  and the upper limits correspond to a 3$\sigma$ level.  $^{d}$  Assuming a \rco=$1.0\pm0.5$. The statistical errors associated to the molecular gas include \rco\ uncertainties.}
\end{table*}

\section{LBQS0109}
\label{sec:lbqs}

\begin{figure*}
\centering
{\LARGE  (a) }
\includegraphics[width=0.7\columnwidth]{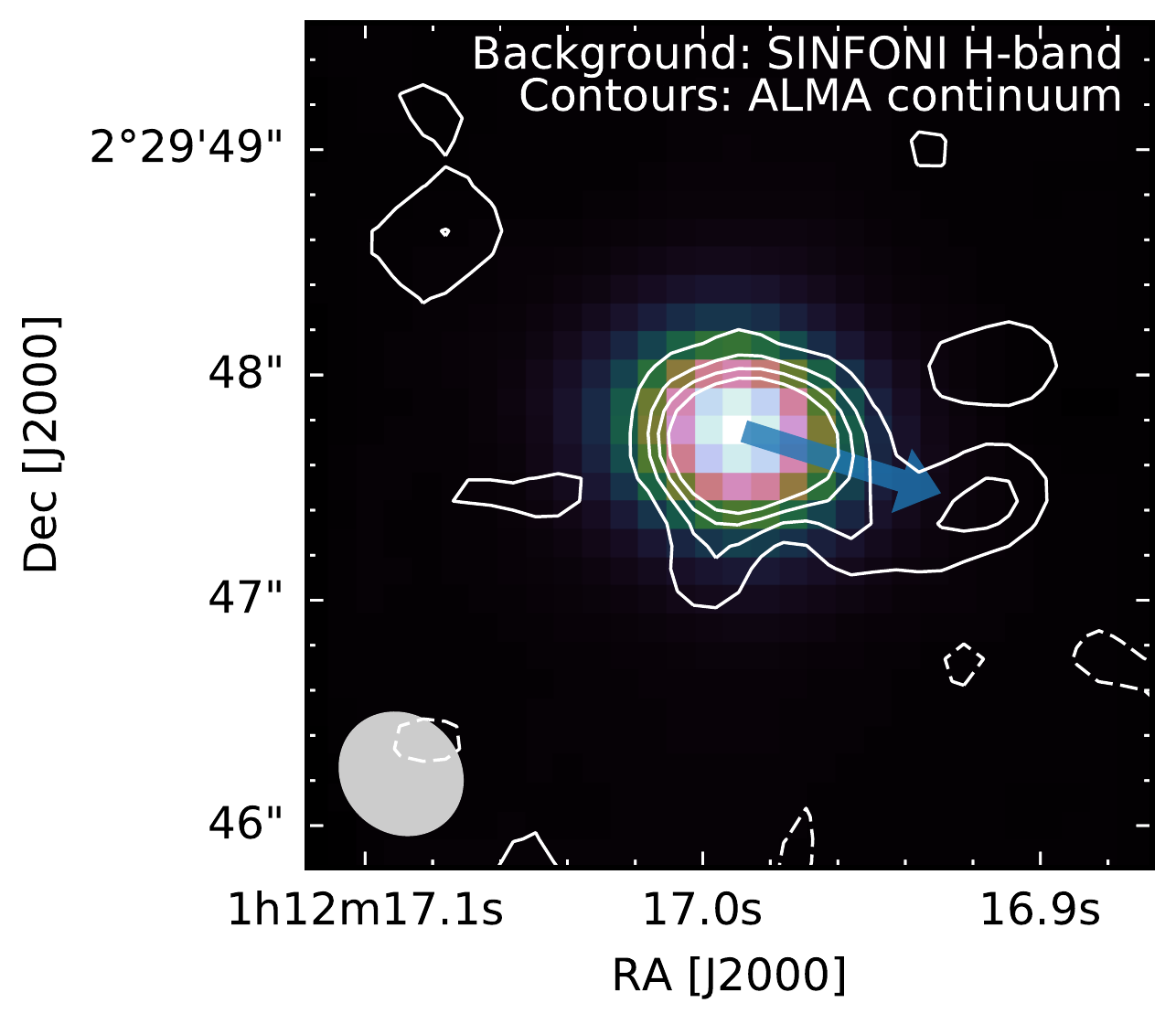}\\\ \\
\ \\ 
 {\LARGE  (b) }
  \includegraphics[width=0.8\columnwidth]{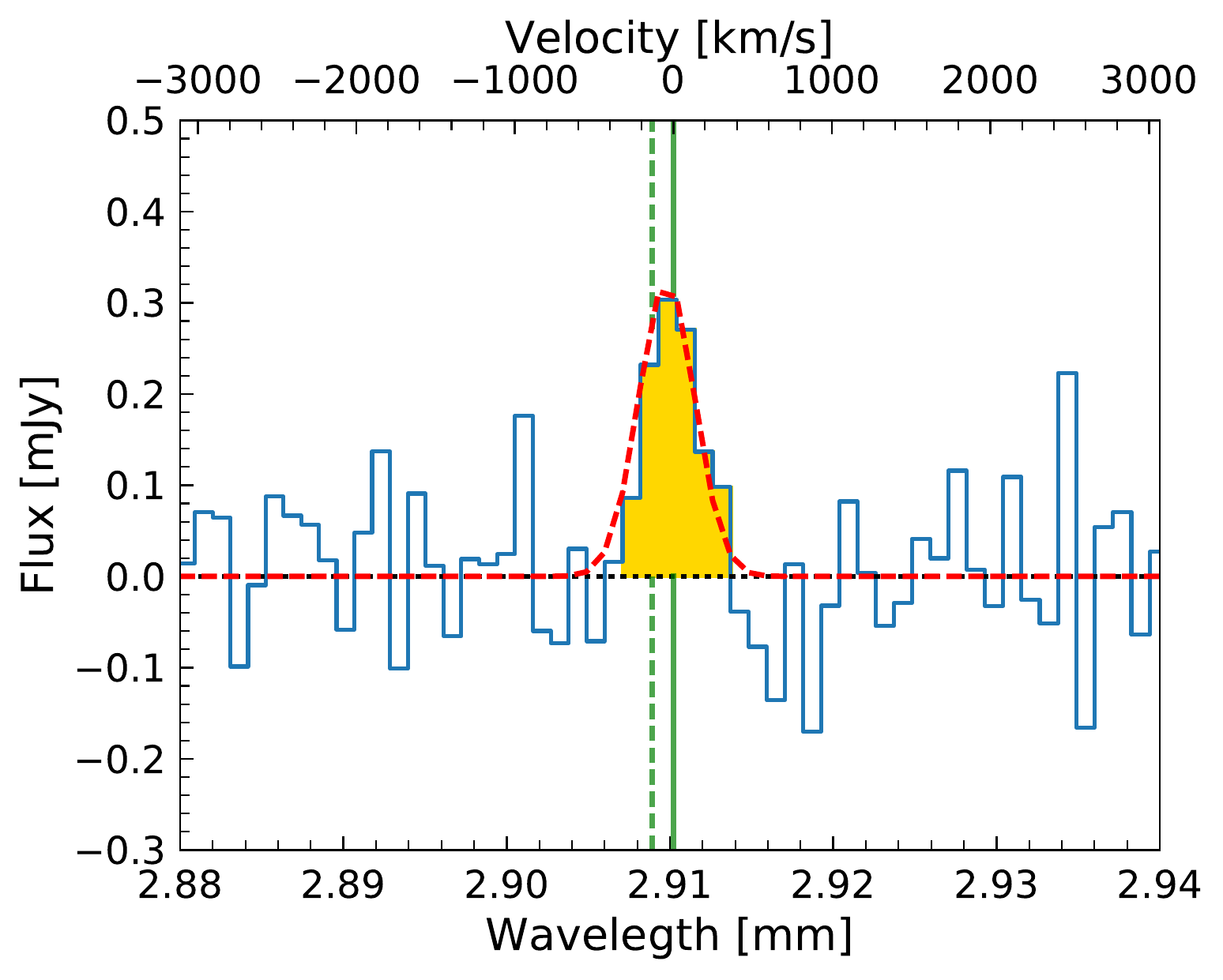}  \quad \quad  {\LARGE  (c) }
  \includegraphics[width=0.8\columnwidth]{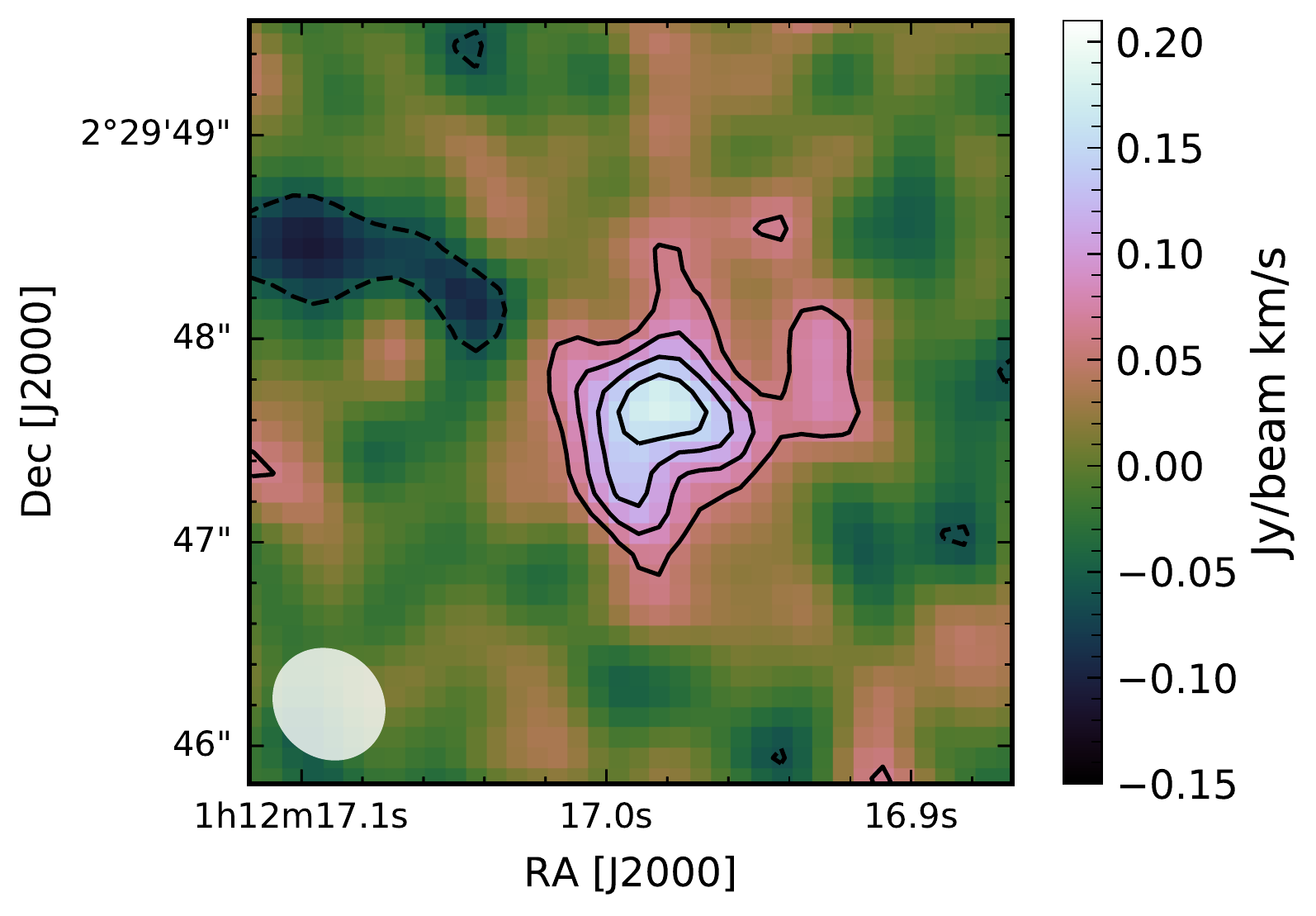}
 \caption{\lb\ : (a) White contours show the  ALMA 3~mm continuum emission in correspondence of \lb\ at an angular resolution of 0.64\arcsec$\times$0.53\arcsec.  Contours correspond to -2, 2, 3, 4 and 5 times the noise per beam (12 \ujy). The colour background image shows  the continuum emission in H band from SINFONI observations. The blue arrow indicates the direction of the ionised outflow revealed by the kinematic analysis of the broad \oiii\ line \citep{Carniani:2015}.
 (b)  CO(3-2) spectrum extracted from an aperture as large as the  synthesised beam size of the ALMA observations and re-binned to 120 km/s. The  vertical green solid and dotted  lines mark the expected  CO(3-2) central wavelength based on the redshift of the narrow \ha\ and \oiii\ component, respectively \citep{Carniani:2016}. The dashed red curve shows the best fit Gaussian profile.
  (c) CO(3-2) surface brightness map. Black solid contours are at the levels of 2$\sigma$,3$\sigma$,4$\sigma$ and 5$\sigma$ of the CO(3-2) flux map, where $\sigma$ is 0.03 Jy/beam km/s.  The $2\sigma$ negative contours are indicated by the black dashed curves. The synthesised beam is shown in the bottom-left corner of the map. 
  }
 \label{fig:lbqs0109}
\end{figure*}

The  ALMA 3~mm continuum emission map of \lb\ is shown in  panel (a) of Fig.~\ref{fig:lbqs0109} overlaid onto the SINFONI H-band  continuum image. 
\lb\ has a flux density of $165$ \ujy\ and is detected  at the $\sim14\sigma$ level ($\sigma=12$ \ujy) in the ALMA continuum map.

The emission at 3 mm is spatially resolved with a beam-deconvolved size of $(0.5\pm0.1)\arcsec\times(0.3\pm0.2)\arcsec$ with P.A. = $(85\pm11)\degr$ (Table~\ref{tab:alma}).
We note that the emission is elongated in the same direction (east-west) of the ionised outflow traced by the broad \oiii\  component \citep{Carniani:2015a}, which is indicated as a blue arrow in  Fig.~\ref{fig:lbqs0109}a.

The radio emission of $\log_{10}(L_{8.4}/{\rm W \ Hz^{-1}}) = 24.83\pm0.22$ at 8.4 GHz  \citep{Hooper:1995} measured with the Very Large Array indicates that \lb\ is just below the limit to be classified  as radio-loud QSO ($\log_{10}(L_{8.4}/{\rm W \ Hz^{-1}})>25$).
Therefore the elongated continuum emission at 3 mm may be associated with a radio jet co-spatial with the ionised outflow. 
However, two photometric measurements at 3 mm, which corresponds to  $\lambda_{\rm rest}\sim0.9$ mm in the rest frame,  and at 8.4 GHz ($\lambda_{\rm rest}\sim 10$ mm), are not sufficient to perform a spectral energy distribution (SED) fitting decomposition. 
Indeed a typical  galaxy SED from radio-to-infrared wavelengths can be modelled as a linear sum of dust continuum, thermal bremsstrahlung  and synchrotron emission \citep[see in detail][]{Yun:2002} but we would need more photometric data to disentangle the various components.
Therefore it is not possible to establish whether dust or synchrotron emission dominates the sub-mm continuum in \lb.

Assuming that the continuum emission at 3 mm is mainly associated with dust thermal emission, we estimated an upper limit on the dust content of \lb. 
 Since dust is  rarely optically thick at millimetre wavelengths  a part from a few extreme starburst galaxies \citep[e.g.][]{Soifer:1999, Klaas:2001, Matsushita:2009}, we have adopted the optically thin approximation for our unobscured QSO.
The total dust mass  is thus given by:
\begin{equation}
\centering
M_{\rm dust} = \frac{S_\nu D_{\rm L}^2}{B_\nu(T_d)\kappa_\nu },
\end{equation} 
where $S_\nu$ is the flux density at the rest frame frequency $\nu$, $D_{\rm L}$ is the luminosity distance of the target, $B_\nu(T_d)$ is the black-body function at the dust temperature $T_d$ and $\kappa_\nu$ is the  absorption coefficient. 
Following \cite{Palau:2013}, the equation can be simply rewritten as 
\begin{equation}
M_{\rm dust} = 3.25\times10^9 \ \msun\ \frac{e^{0.048\nu/T_d}-1}{\nu^3\kappa_\nu}\left(\frac{S_\nu}{\rm mJy}\right)\left(\frac{D_L}{\rm Mpc}\right)^2,
\end{equation} 
where the rest frame frequency $\nu$ is in GHz, $\kappa_\nu$ is in ${\rm cm^2 g^{-1}}$ and $T_d$ is in K.
We adopt a $\kappa_\nu$ = ${\rm 0.45\times(\nu/250 GHz)^\beta \ cm^2 g^{-1}}$ with a fixed emissivity index $\beta=2.0$. 
We note that $\kappa_\nu$ depends on the properties of dust grain and can suffer from large uncertainties.
Using a dust temperature $T_d=40-60 K$ \citep{Beelen:2006}, the  dust mass from the 3 mm continuum emission is around $5-8\times10^{8}$ \msun.
We stress that the inferred dust mass is more likely an upper limit because a fraction of the continuum emission in this wavelength band can be also associated to non-thermal synchrotron radiation.

In addition, the ALMA continuum map of the field surrounding \lb\  reveals the presence of three serendipitous sources within 15\arcsec\ ($\sim120$ kpc) from the QSO and with a signal-to-noise ratio (S/N)$>4$.
The three sources are also visible with a level of confidence $> 5\sigma$ in the line  channels. 
By assuming that the  line emission detected  at the positions of these sources corresponds to the CO(3-2) transition, we propose that they are physically associated with \lb, with a $\Delta z$=0.002.
The properties inferred from the analysis of these serendipitous detections are reported in Appendix~\ref{sec:appA}. 
The three detections do not have optical counterparts in the SDSS images. A detailed discussion of this over-density system is presented in Sect.~\ref{sec:overdensity}.


The CO(3-2) spectrum of \lb\ is shown in Fig.~\ref{fig:lbqs0109}b, and the vertical green  dotted line shows the expected central wavelength corresponding to the redshift measured from the SINFONI data by using the narrow \oiii\ emission.
The  full-width at half maximum (FWHM) and centroid  of the CO(3-2) line in \lb, derived from a single Gaussian fit, are $400\pm60$ km/s and $\lambda_{\rm obs}=2.9094\pm0.0004$ mm ($z_{\rm CO(3-2)} = 2.3558\pm0.0005$).
We note that {both the FWHM and the redshift} are consistent  with the narrow \oiii\ and \ha\ components (FWHM$_{\rm \sc [OIII]} = 490\pm90$ km/s,  FWHM$_{\rm H\alpha} = 250\pm200$ km/s, $z_{\rm \sc [OIII]}=2.3558\pm0.0008$, and $z_{\rm H\alpha}=2.357\pm0.002$) 
tracing SF in the host galaxy \citep{Carniani:2016}. 
 The agreement strongly indicates that most of the CO emission is tracing molecular gas in the host galaxy.

Panel (c) of Fig.~\ref{fig:lbqs0109} shows the map integrated over the line emission in which the peak is detected at $\sim7\sigma$ ($\sigma=0.03$ Jy/beam km/s).
The integrated flux, extracted from the region of the map with a level of confidence higher than 2$\sigma$, is ${\rm S_{CO(3-2)}\Delta v = 0.34\pm0.03}$ Jy km/s, corresponding to a line luminosity of ${\rm L_{CO(3-2)} = (1.4\pm0.2)\times10^7}$ \lsun\ at $z=2.35$.
The CO line emission is spatially resolved  with an estimated size of $(0.93\pm0.08)\arcsec\times(0.82\pm0.07)\arcsec$, that is $(7.7\pm0.7)\ {\rm kpc}\ \times(6.8\pm0.6)$ kpc.

Although the CO(3-2) line emission is resolved by our ALMA observations, we cannot perform a detailed pixel-by-pixel kinematic analysis  because of the  low S/N  of the data. However, we note that the CO line extracted in the southern region has a  FWHM= 280 km/s that is smaller than that measured in the nuclear region (see Table 1). 
This discrepancy will be discussed in more detail in Sect.~\ref{co_lbqs}.

\begin{figure}
  \includegraphics[width=1\columnwidth]{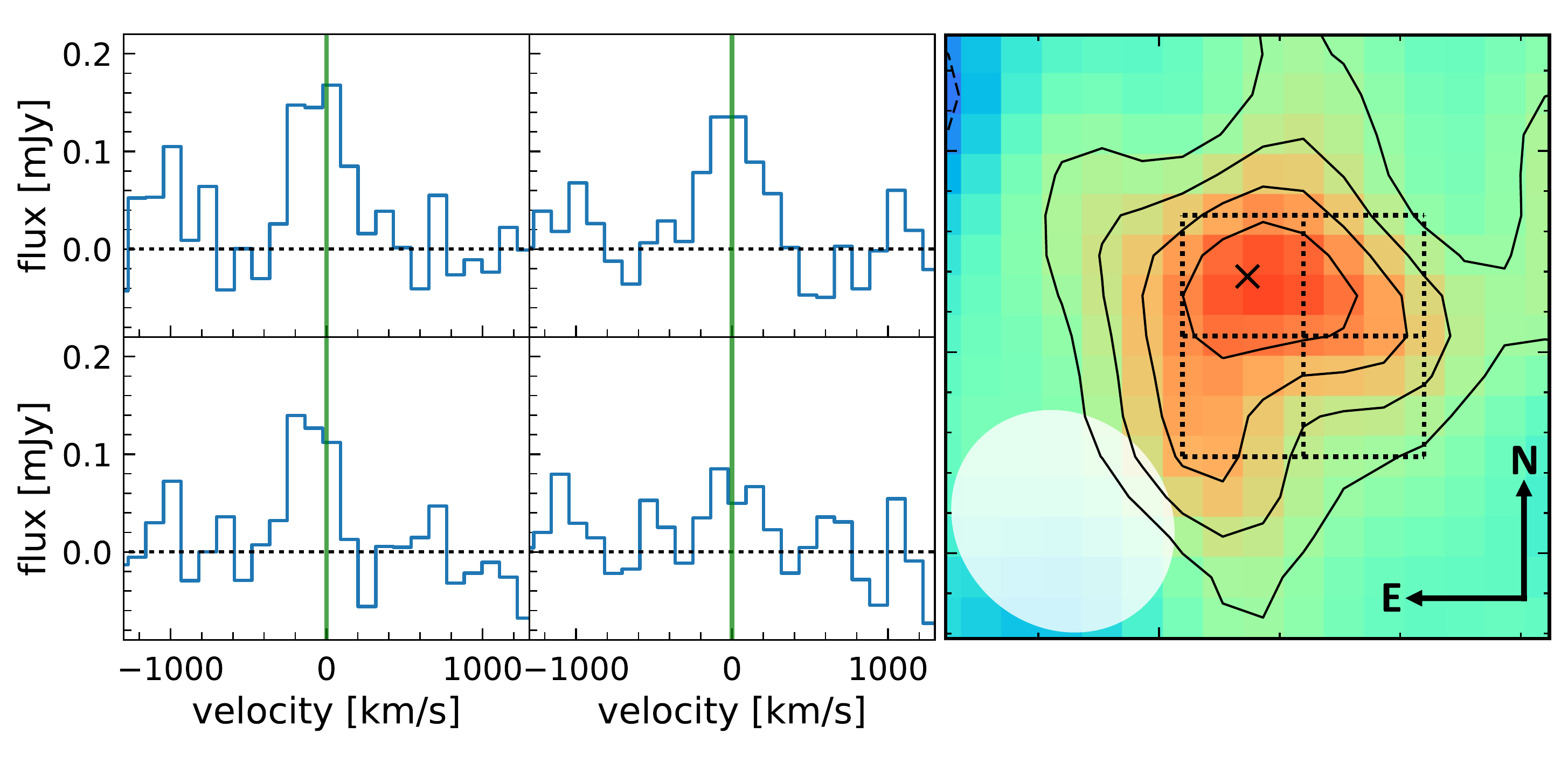}

\caption{ Left: CO(3-2) spectra extracted from four regions (shown in right panel) placed at different positions with respect to the location of the QSO.  The spectra have been rebinned to a channel size of 120 km/s. The CO line emission is faint or absent in the  lower-right spectrum which was extracted from a region south-west of the QSO. Right: CO(3-2) map and contours at the same level of Fig.~1c. The black dotted squares  correspond to the regions  from where we extracted the four spectra shown in the left panel. 
The beam of the CO(3-2) map is shown in the bottom-right corner.} 

 \label{fig:grid}
\end{figure}

\subsection{Molecular and stellar mass estimates}
\label{sec:estimation}

Consistently with recent  high-redshift ($z>2$) observations of   AGN host galaxies \citep{Sharon:2016}, we assume  an ${\rm r_{31} \equiv L^{\prime}_{CO(3-2)}/L^{\prime}_{CO(1-0)}}$ ratio of,  \rco$=1.0\pm0.5$, yielding an estimated CO(1-0) line luminosity of, $L^{\prime}_{\rm CO(1-0)} = (1.0\pm0.6)\times10^{10}$ K km/s pc$^2$ for \lb.

The conversion factor \aco\ between CO(1-0) line luminosity and H$_2$ mass depends on the interstellar medium conditions. 
In general an \aco=4 \msun/K km/s pc$^2$ is assumed for main-sequence (MS) galaxies and an \aco=0.8  \msun/K km/s pc$^2$ value is adopted for compact luminous systems, such as starburst galaxies, SMGs and QSOs \citep{Downes:1998, Carilli:2013, Bolatto:2013}.
 Since the molecular ISM conditions of our targets are still unknown, we  estimate two limiting values for the molecular gas mass (\mgas),  corresponding to the two  \aco\ choices mentioned above. The  resulting \mgas\ values are listed in Table~\ref{tab:alma} and  their associated statistical errors  include both ALMA flux calibration and  \rco\ uncertainties. 

We now explore the consequences of the  possibility that the host galaxy of \lb\ lies on the  main sequence (MS) of star forming galaxies. Typical MS galaxies at $z\sim2$ have molecular gas fractions of $f_{\rm mol-gas}$=\mgas/(\mgas+\mstar)$\simeq$ 0.44 \citep{Tacconi:2010}.  For \lb\ and assuming \aco=4 \msun/ km/s pc$^2$, this gas fraction would result in a stellar mass estimate of, \mstar=1.3$\times10^{11}$ \msun. This \mstar value, combined with the BH mass of \mbh=$10^{10}$ \msun\ inferred by \cite{Shemmer:2004}, yields for \lb\  \mbh/\mstar$\simeq0.1$.
This ratio is much larger than those observed in massive galaxies in the local Universe \citep{Kormendy:2013}.
A similar \mbh/\mstar\ has been recently  inferred by \cite{Trakhtenbrot:2015} for an unobscured AGN at z = 3.328, CID947,  where it is believed that the SMBH has grown more efficiently than the host galaxy.  
 In \lb\ the star-formation activity may have been shut-off due to the negative-feedback exerted by the QSO, as we further argue below. It is also possible of course that the host galaxy of \lb\ is not on the $z\sim2$ MS, in which case the above estimate of \mbh/\mstar\ would not be valid.

\subsection{Morphology of the CO(3-2) emission}
\label{co_lbqs}

%

\begin{figure*}  
\includegraphics[width=0.65\columnwidth]{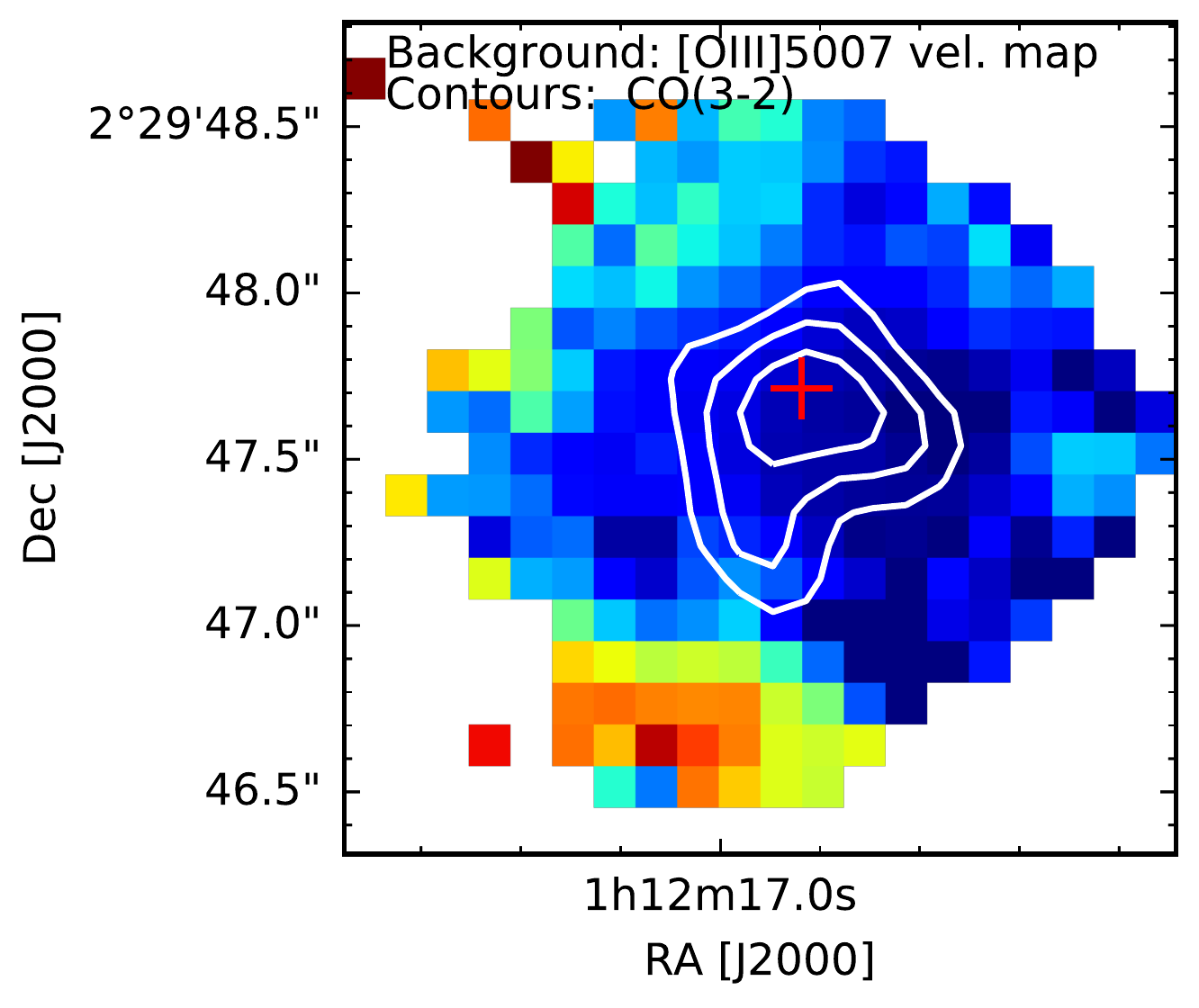}
 \includegraphics[width=0.65\columnwidth]{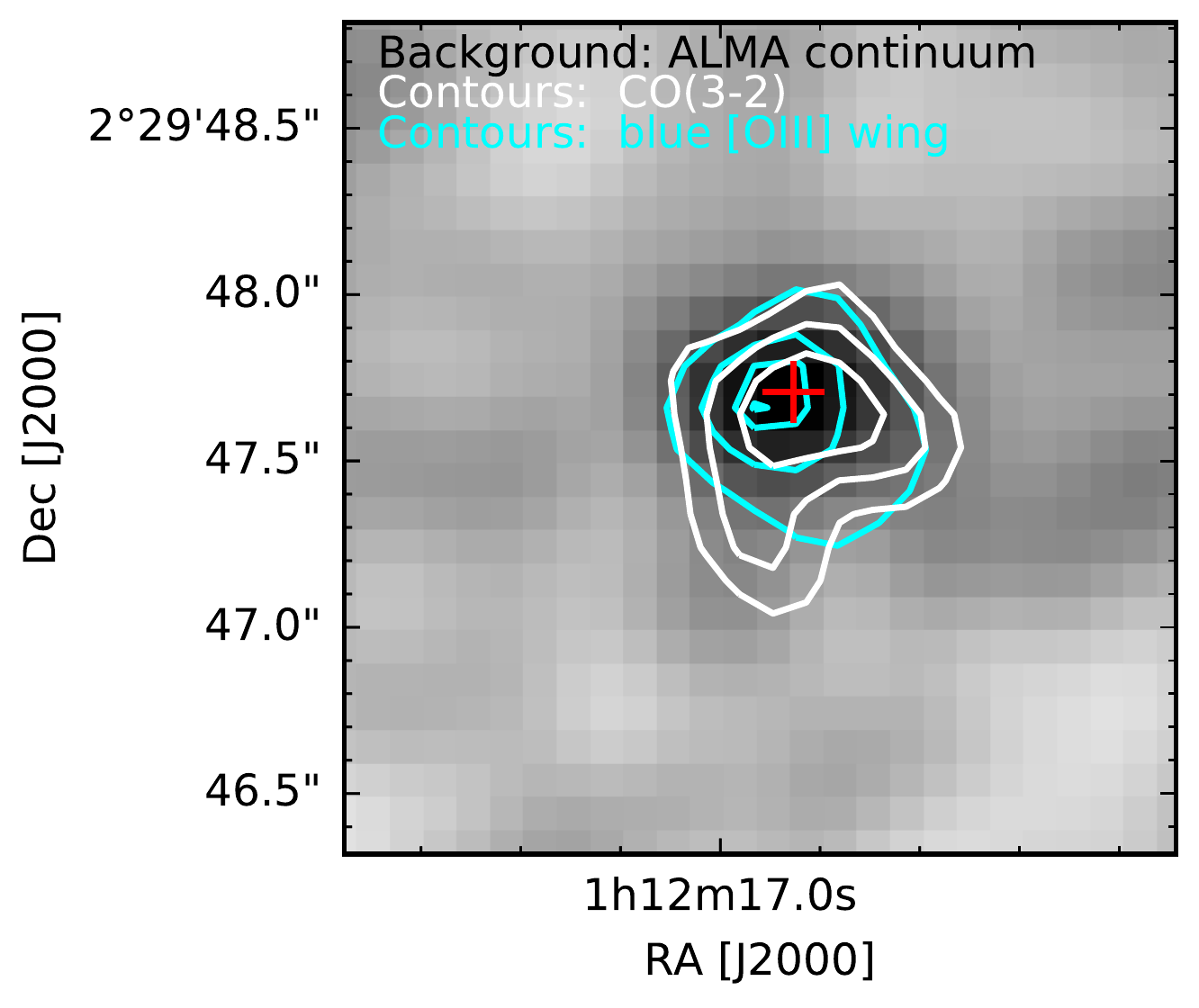}
 \includegraphics[width=0.65\columnwidth]{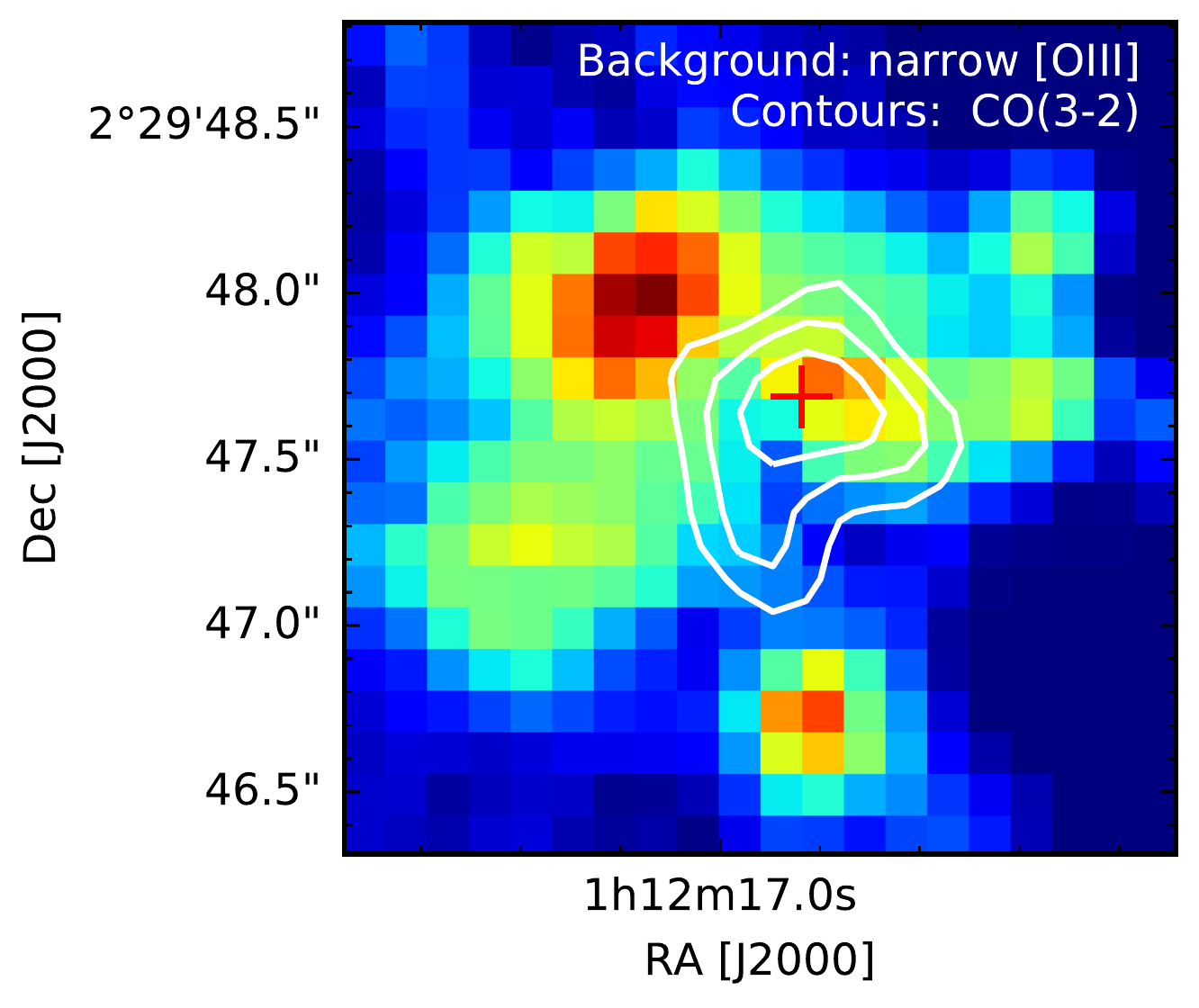}
\caption{  \lb: from left to right:  \oiii\ velocity map by Carniani et al. 2015b, ALMA continuum map at 3 mm, and narrow [OIII] emission tracing SF in the host galaxy \citep{Carniani:2016}.  
White contours trace the flux map  of CO(3-2) emission at the  levels 3, 4 and 5$\sigma$. The red cross indicates the centroid of the H-band continuum emission. The cyan contours in the middle panel shows  the flux map of the blue \oiii\ wings collapsing the SINFONI cube over the velocity range  $-1750<v<-1950$ km/s. 
} 
 \label{fig:oiii_co}
\end{figure*}

The  CO(3-2) map of \lb\ shown in Fig.~\ref{fig:lbqs0109}c exhibits a complex morphology: the molecular emission is not  distributed symmetrically around the QSO. 
 Figure~\ref{fig:grid} shows the CO(3-2) spectra extracted from four regions placed at different positions with respect to the location of the QSO. 
The CO(3-2) emission in the south-west region  is almost absent, while the line spectra extracted from  the other three regions have similar fluxes and profiles.  We note that the synthesised beam is oriented from north-east to south-west. Thus, the signal visible towards the west and south relative to centre suggests that the molecular emission the molecular emission is either spatially unresolved or faint  along the direction of the ALMA beam.

In the left panel of Fig.~\ref{fig:oiii_co} we compare the  distribution of CO(3-2) with the velocity map of the broad  blue-shifted \oiii\ component tracing the ionised outflow in the QSO host galaxy \citep{Carniani:2015a,Carniani:2016}. 
 The  CO(3-2)  emission  is partially dislocated with respect to the regions where the outflow traced by \oiii\ is fastest.   
In addition, the \oiii\ channel map, obtained integrating the continuum subtracted SINFONI datacube on  the blue wing, indicates that the ionised outflow is elongated from north-east to south-west (cyan contours in the middle panel) where the CO(3-2) emission is faint (or spatially unresolved).
Finally, the right panel of Fig.~\ref{fig:oiii_co} shows the surface brightness  of the narrow \oiii\ component tracing SF in the \lb\  host galaxy and the CO(3-2) flux map in white contours. 
We refer to \cite{Carniani:2016} for further arguments supporting the identification of the narrow \oiii\ emission with emission powered by star formation in the quasar host galaxy.
It is interesting to note a similarity between the CO and the narrow \oiii\ surface brightness distributions.
Both emission lines are faint or absent  along the direction of the ionised outflow, while they are clearly visible in the other regions. 
These results support a scenario in which fast outflows are cleaning up the galaxy of its molecular  gas, hence quenching SF in  the region where the outflow breaks in the host galaxy ISM. 

In Sect. \ref{sec:lbqs} we noted that the CO(3-2) profile extracted from an aperture  placed south of the QSO (see Fig.~\ref{fig:lbqs0109}) is narrower than that observed in the nuclear region. 
Such a discrepancy suggests that the motion of the gas in the external regions is less turbulent  than in the QSO centre which is influenced by the nuclear fast winds. 
This residual gas fuels the SF in the region of the host galaxy is not affected by AGN-driven outflows. 

The current ALMA CO(3-2) observations trace  molecular gas only in region within 2 kpc from the centre, while the narrow \oiii\ component is extended up to $\sim8$ kpc.
Unfortunately, higher sensitivity ALMA observations would be needed to compare the distribution of molecular gas and \oiii\ emission in the external regions  at a distance $>2$~kpc from the QSO, and to verify whether the extended structures are  consistent with the Schmidt-Kennicutt relation between SFR and gas density \citep{Kennicutt:1998}.

\section{2QZJ0028}
\label{sec:2qzj0028}

\begin{figure*}
\centering
{\LARGE  (a) }
\includegraphics[width=0.7\columnwidth]{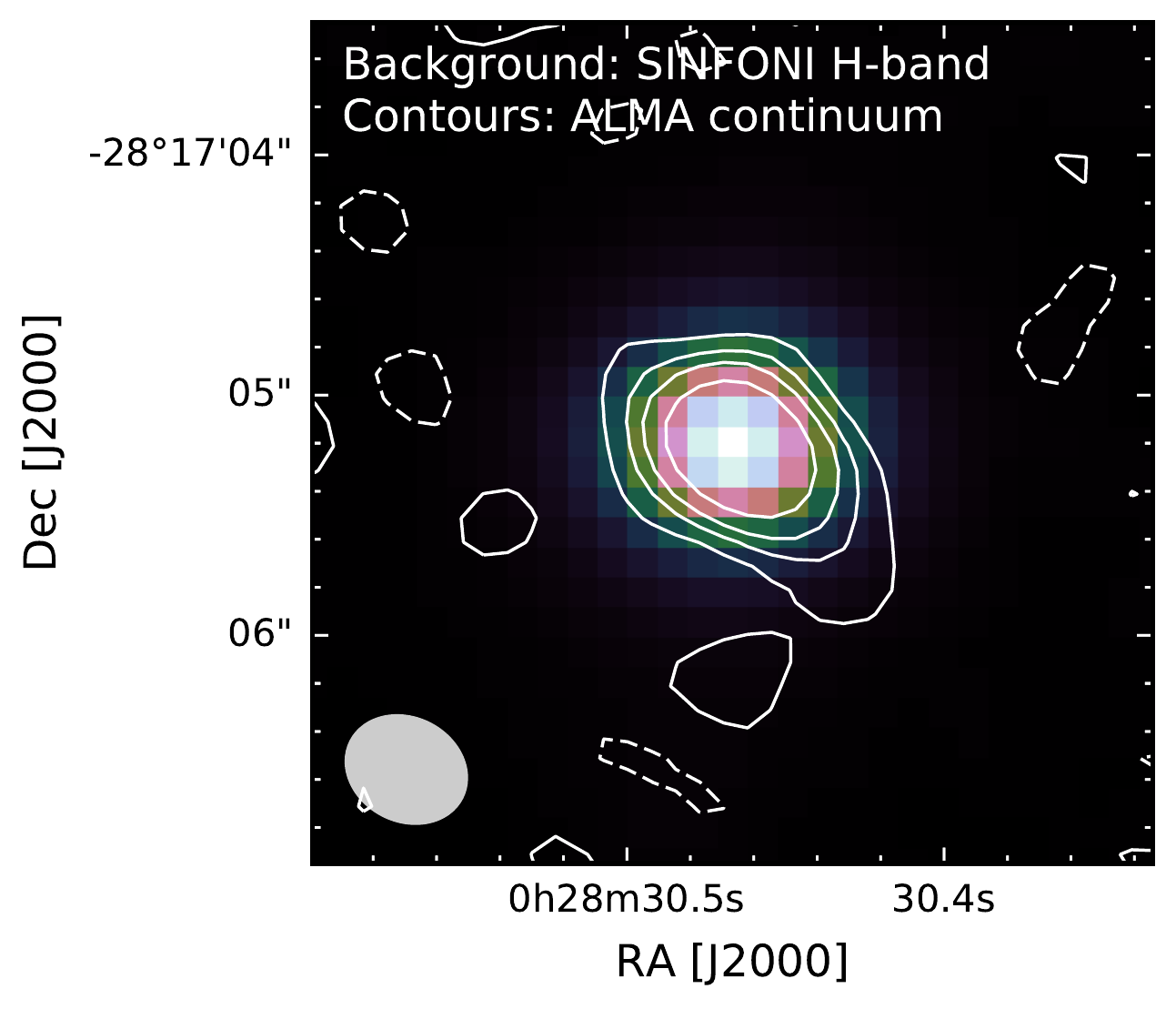}\\
\ \\
\ \\ 
 {\LARGE  (b) }
  \includegraphics[width=0.8\columnwidth]{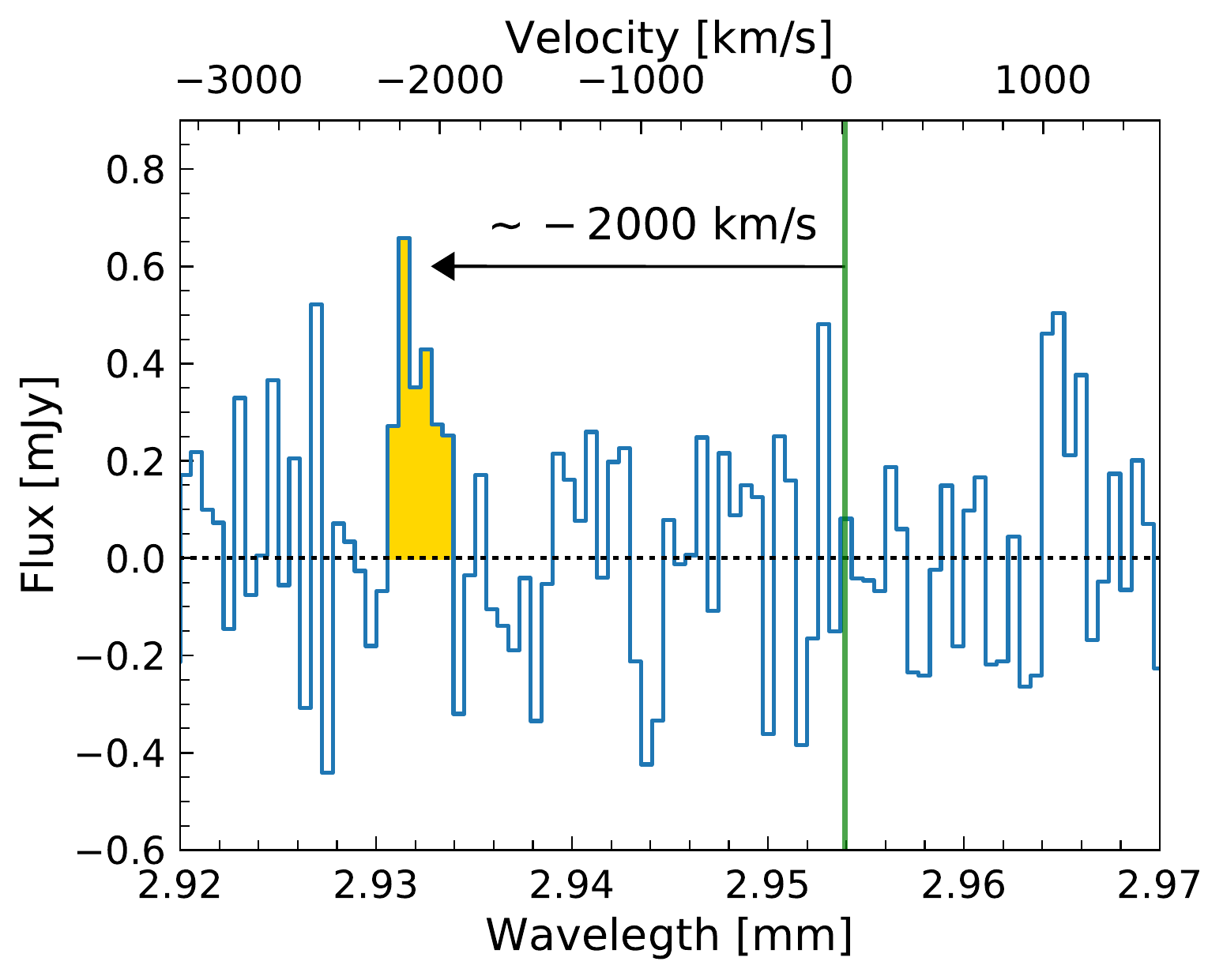}
  \quad \quad  {\LARGE  (c) }
  \includegraphics[width=0.8\columnwidth]{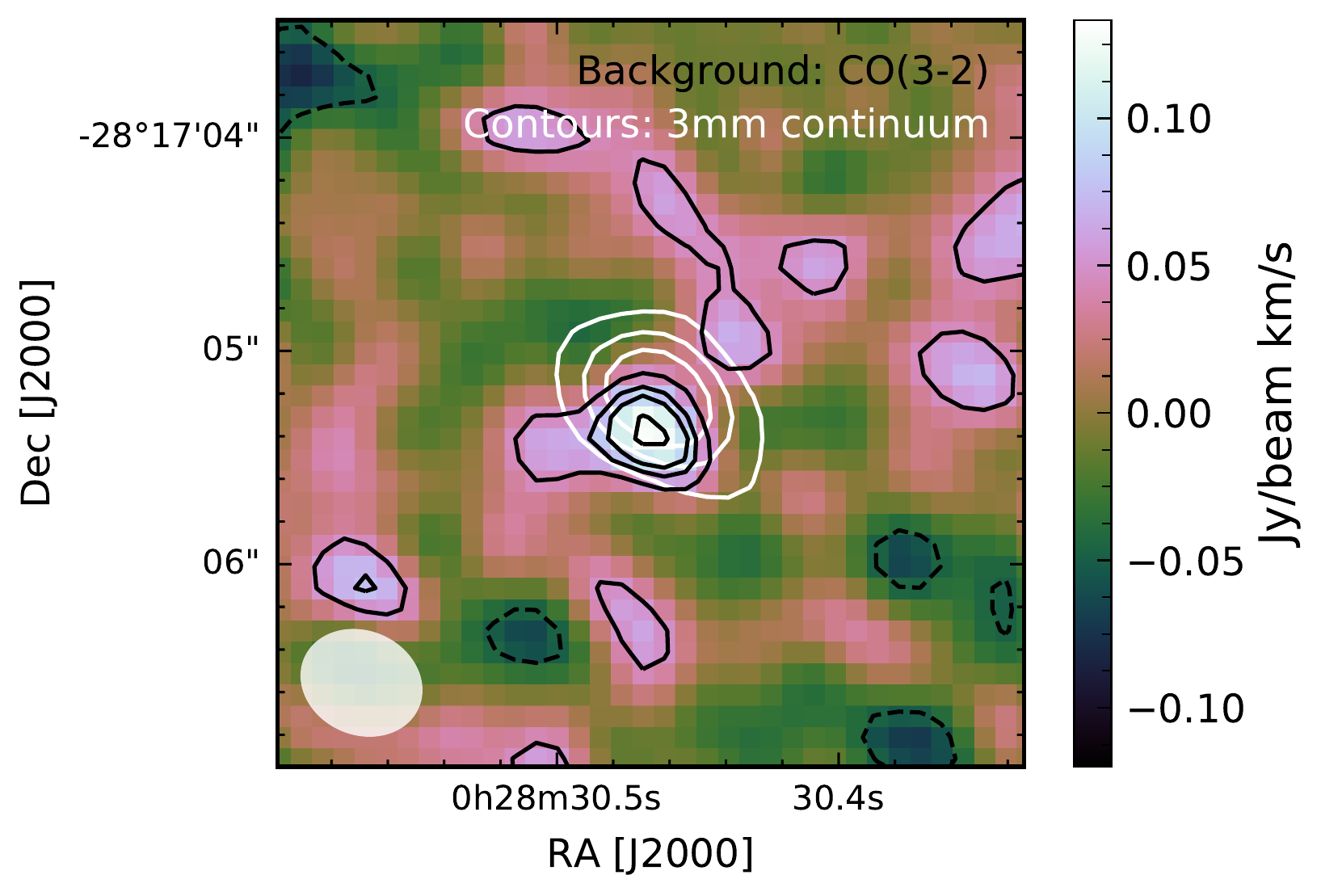}
 \caption{ \qz\ : (a) White contours show the continuum emission from \qz\ at 3 mm with a bean of 0.64\arcsec$\times$0.53\arcsec. Contours are at the level of 3, 5 and 7 times the noise per beam (12 \ujy). The color background image shows the continuum emission in H-band from SINFONI data of \qz.
 (b)  ALMA spectrum  extracted from an aperture as large as the beam size and rebinned to 60 km/s. The vertical dashed green line  mark the expected positions for CO(3-2) emission line at the redshifted of the narrow \ha\ component \citep{Cano-Diaz:2012}: the line is not detected. An emission line is detected at a velocity of $\sim-2000$ km/s with respect to the redshift of 2QZJ0028.
 (c) CO(3-2) surface brightness. Black solid contours are at the levels of 2$\sigma$, 3$\sigma$, 4$\sigma$ and 5$\sigma$, where $\sigma$ is 0.03 Jy/beam km/s. The white contours show the continuum emission at 3 mm at the levels of 3, 6, and 9 times the sensitivity of the continuum map. The synthesised beam is shown in the bottom-left corner.} 
\label{fig:2qzj0028}
\end{figure*}

 Panel (a) of Fig.~\ref{fig:2qzj0028} shows in white contours the spatially unresolved continuum emission map of \qz at 3~mm, while the coloured background is the SINFONI continuum emission in the H band. The peak at 3 mm  has a  S/N=14 and the integrated flux density of the source is 170$\pm12$ \ujy.
 
We cannot perform a radio-to-FIR SED fitting decomposition because we have only one photometric point. 
However, by assuming that the emission at 3 mm is mainly associated to dust thermal emission we can infer the dust mass as we did for \lb\ in Sect.~\ref{sec:lbqs}.
We thus estimate M$_{\rm dust} = 6-9\times10^{8}$~\msun

 We have performed a blind search in the ALMA continuum map around the QSO and we have detected a millimetre continuum source with a  confidence level of $5\sigma$ and a flux density of 61 \ujy\ at 3 mm. Such source is located at a distance of 118 kpc from the quasar (Appendix~\ref{sec:appA}).

At the redshift of  \qz\, we do not detect any CO(3-2) emission line at a significance level higher than $3\sigma$ (panel (b) Fig.~\ref{fig:2qzj0028}). By assuming a CO line width similar to that measured in \lb\ (FWHM=400 km/s), we can estimate a 3$\sigma$ upper limit on the CO(3-2) integrated line flux of 0.09 Jy km/s, which corresponds to an upper limit on the CO(3-2) line luminosity of $0.6\times10^7$ \lsun.
Following the same method as in Sect.~\ref{sec:estimation},  we derive two different $3\sigma$ upper limit estimates for the total molecular gas mass, based on different \aco\ prescriptions: 
M$_{\rm CO}$(\aco=0.8)  $<0.2\times 10^{10}$ \msun\ and 
M$_{\rm CO}$(\aco=4) $< 1.2\times 10^{10}$ \msun.

\subsection{Possible association of offset CO(3-2) emission with a molecular outflow}
\label{sec:possible_detection}

\begin{figure*}  
\includegraphics[width=0.65\columnwidth]{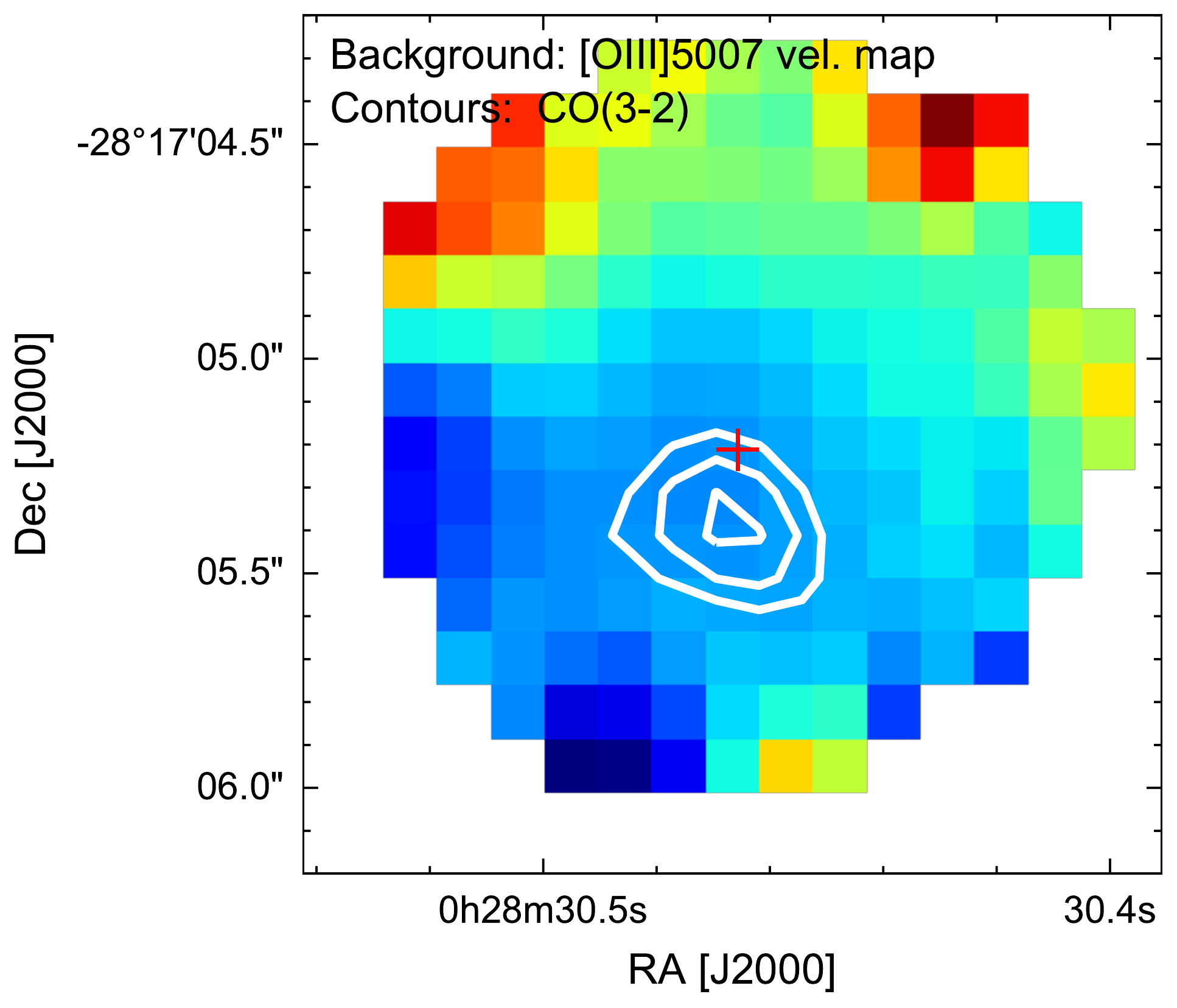}
 \includegraphics[width=0.65\columnwidth]{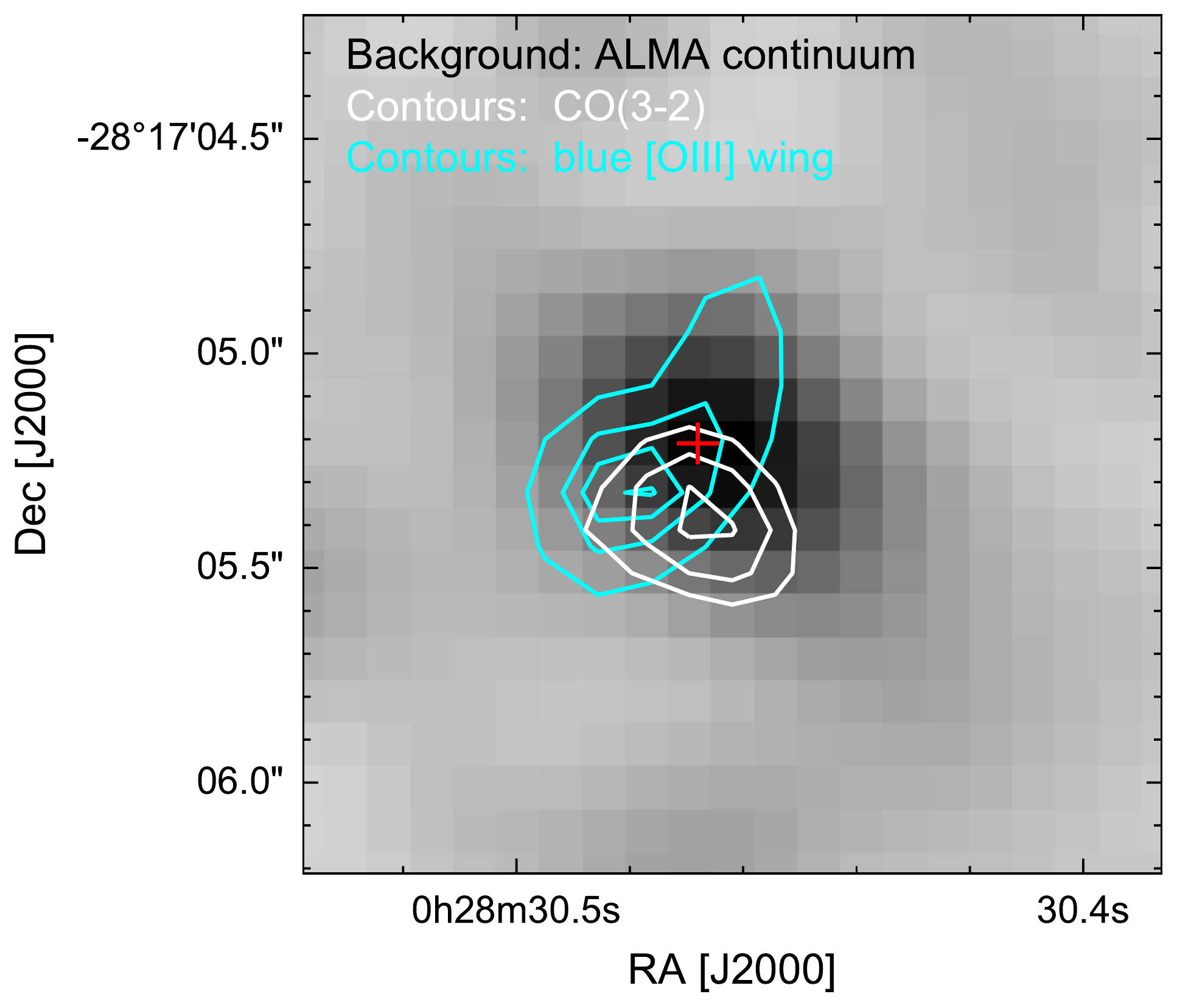}
 \includegraphics[width=0.65\columnwidth]{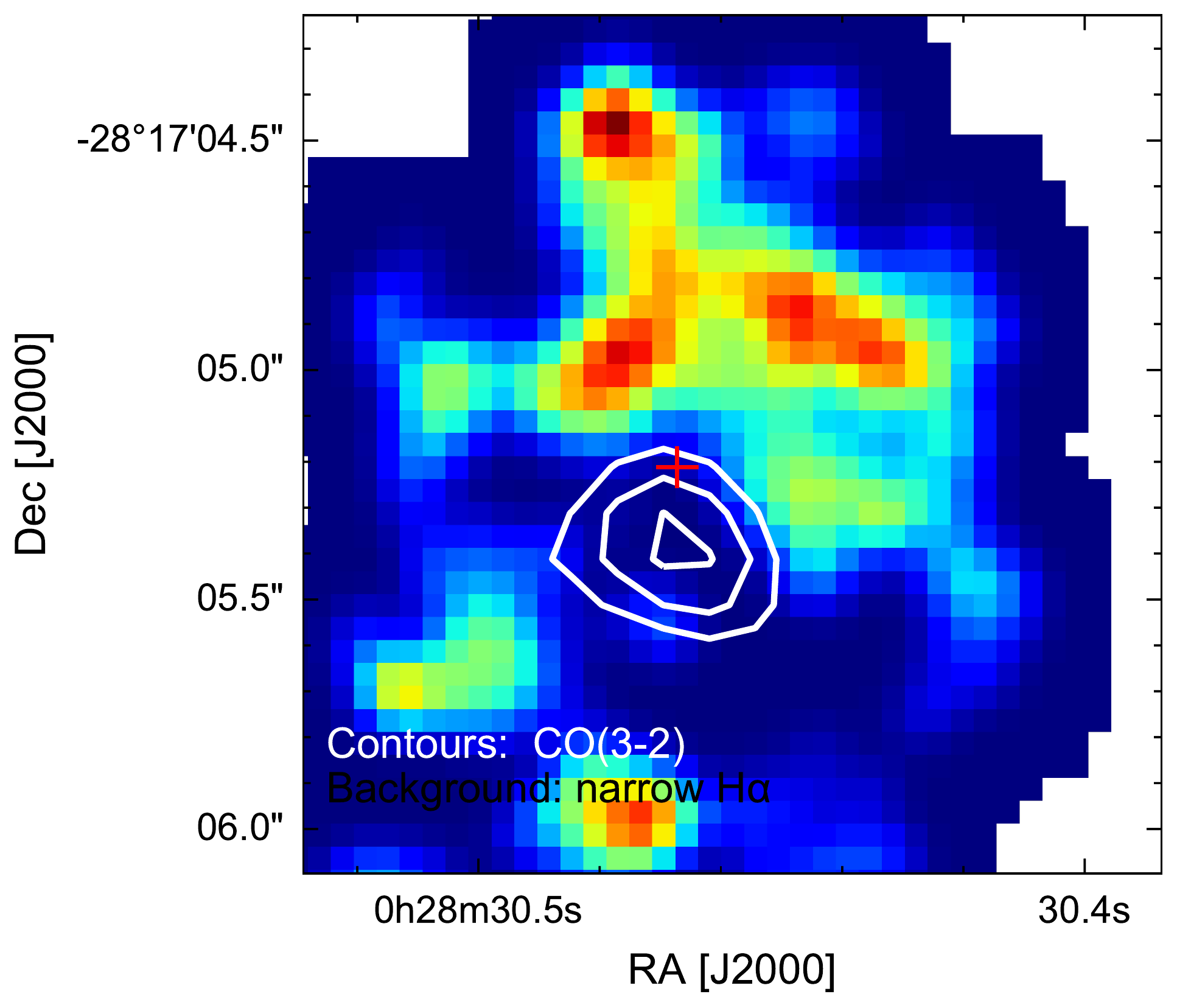}
\caption{ \qz: from left to right:  \oiii\ velocity map by Carniani et al. 2015b, ALMA continuum map at 3 mm, and narrow \ha\ emission tracing SF in the host galaxy \citep{Cano-Diaz:2012}.  
White contours trace the flux map pf CO(3-2) emission at the same levels of t the  levels 3, 4 and 5$\sigma$. The red cross indicates the centroid of the H-band continuum emission. The cyan contours in the middle panel shows  the flux map of the blue \oiii\ wings collapsing the SINFONI cube over the velocity range  $-2500<v<-2300$ km/s. 
} 
 \label{fig:sinfoni_ALMA_CO}
\end{figure*}

The total spectrum extracted at the location of \qz\ from a beam-sized aperture shows an emission feature at a velocity of $\sim-2000$ km/s relative to the  expected CO(3-2)  frequency based on the QSO redshift (Fig.~\ref{fig:2qzj0028}b).
Figure~\ref{fig:2qzj0028} (panel c) shows the map extracted from the spectral range centred at 2.9316 mm ($\sim-2000$ km/s) and with a spectral width of 250 km/s. This map clearly shows an unresolved source whose peak is detected with S/N=5.2 and is spatially offset by $\sim0.2\arcsec$ (1.3 kpc) towards south-east relative to the QSO centre. 
Although we cannot completely rule out that this detection is spurious, we note that (as discussed in Appendix A) the number of positive peaks  at $>5\sigma$ is 1.5 times larger than the number of negative peaks, suggesting that one third of the positive peaks might represent real sources.
Similarly, the emission feature at $\sim1000$ km/s should not be considered as real because in the integrated map the emission peak has a S/N of only 3.8; at that S/N the number of positive peaks is similar to that of negative ones, strengthening the idea that it is simply a noise fluctuation.
In any case, deeper observations are needed to confirm the reliability of our CO detection at the velocity of $\sim-2000$ km/s.

\begin{table}
\caption{Properties of the faint blueshifted CO detection in \qz.}           
\label{tab:qz}      
\centering          
\begin{tabular}{l c }    


\hline
\\
$\lambda_{\rm CO(3-2)}$ [mm] &  $2.9316\pm0.0004$  \\
FWHM$_{\rm CO(3-2)}$ [km/s] & $250\pm90$  \\

S$_{\rm CO(3-2)} \Delta v$ [Jy km/s]  & $0.12\pm0.02$  \\

L$^\prime_{\rm CO(3-2)}$ [$10^{10}$ K km/s pc$^2$] & $0.37\pm0.06$  \\

L$_{\rm CO(3-2)}$ [$10^{7}$ \lsun]  & $0.5\pm0.1$  \\


\\
\hline
\\              
\end{tabular}   
\end{table}

Under the assumption that such detection is real and associated to the CO(3-2) transistion, Table~\ref{tab:qz} summarises the properties of the line.
Both  the central velocity and the positional offset of the  blue-shifted CO feature are consistent with the velocity ($\sim2300$ km/s) and location of the ionised outflow traced  by the broad \oiii\ emission.  This is clearly shown in left panel of Fig.~\ref{fig:sinfoni_ALMA_CO} where we compare the flux map of this tentative CO(3-2) component with the ionised gas velocity map obtained from the SINFONI observations. Such a remarkable agreement strongly suggests that the blue-shifted CO(3-2) component detected by ALMA is real and that it traces a molecular outflow.

The middle panel of Fig.~\ref{fig:sinfoni_ALMA_CO} shows the flux map of the CO(3-2) emission outlined over the ALMA continuum emission (background image).  In this panel, we plot the surface brightness of the broad \oiii\ emission line  obtained by collapsing the SINFONI data-cube in a velocity range $-2500<v<-2300$ km/s, where our spectro-astrometry technique has revealed the presence of an extended outflow \citep{Carniani:2015a}.
The CO blueshifted emission overlaps with the blueshifted \oiii\ emission suggesting that  we may be tracing an outflowing molecular component associated with the ionised outflow. 
The radius of the ionised outflow, estimated by using the spectroastrometry method, is $\sim0.7$ kpc, which is consistent, within astrometric error ($\sim$0.9 kpc),  with the spatial offset ($\sim$1.6 kpc) measured between the line and continuum emission at 3 mm. 

The non detection of CO emission at systemic velocity and the host galaxy position can be explained by the different excitation of the molecular gas in the star formation regions and in the outflow.
For instance, from recent ALMA observations of a jet-driven molecular outflow, \cite{Dasyra:2016} found that the CO(4-3) emission is  more excited  along the jet propagation axis than in the rest of the galaxy disk. If the same excitation ratio describes the case of \qz\ then the emission at the systemic velocity would not be detected with our observations.
Blueshifted CO emission like our own without any counterpart at the systemic velocity has been detected in quasars by \cite{Banerji:2016}.
On the other hand, the non detection of CO emission at the systemic velocity may also indicate that a large fraction of the molecular gas in the host galaxy is accelerated by the AGN-driven outflow and the sensitivity of the current ALMA observations is not sufficient to detect the residual quiescent  gas,  even assuming the same excitation ratio for the outflow and star-formation regions. 
Future deeper ALMA observations of higher and lower rotational CO transitions are fundamental to confirm or rule out the hypothesis that the blue-shifted emission is real and traces molecular outflows in \qz.

Outflowing clumps have already been observed by \cite{Cicone:2015} in a QSO at  $z\sim6.4$ (SDSS J1148+5251).
They find clumps of [CII] emission extended up to $\sim30$ kpc
from the nucleus and with velocities $>$1000 km/s.
In addition, the spectral 
fitting to the [CII] extended emission exhibits the presence of narrow 
($\sigma_v\sim100-200$ km/s) and fast ($v>1000$ km/s) clumps similar to that 
observed in \qz.  
In the local Universe, the presence of outflowing clumps  of molecular gas has been revealed by CO(2-1) and CO(3-2) observations of Markarian 231  \citep[Fig.~1 by][]{Feruglio:2015}, a QSO host and ultra-luminous IR galaxy (ULIRG) in the local Universe, as well as in a few other nearby QSOs \citep[e.g.][]{Cicone:2014}.  In several cases, the CO line profiles show the presence of a blue and red wing composed by several `bumps' with different intensity and velocity. Such profiles may be generated by molecular outflows with a multi-clump morphology.

 The sensitivity of our current ALMA observations is  likely not sufficient to  appreciate both the blue and red wings  of the CO line as observed in local molecular outflows \citep[e.g.][]{Cicone:2012, Feruglio:2010, Feruglio:2015}, but  it allows us to marginally detect only the brightest knot of the  clumpy molecular outflow. 
Since the velocity and the positional offset are consistent with those of the \oiii outflow, we  hypothesise that most of the blueshifted \oiii\ emission is co-spatial with  the molecular outflow clump.
We also note that the CO emission is located in the region where the \ha\ emission is missing (right panel of Fig.~\ref{fig:sinfoni_ALMA_CO}).
Such an anti-correlation supports the notion that the AGN-driven outflow  drives gas out of the galaxy and exhausts the fuel necessary to SF.

An alternative interpretation to the outflow scenario could be that the detected blue-shifted CO emission is associated with a merging companion galaxy. However, our SINFONI observations \citep{Cano-Diaz:2012, Carniani:2015a}  do not show any merging signature.
Additional data are  required to further dismiss or validate this possibility.

\section{HB8903}
\label{sec:hb8903}

\begin{figure*}
\centering
{\LARGE  (a) }
\includegraphics[width=0.7\columnwidth]{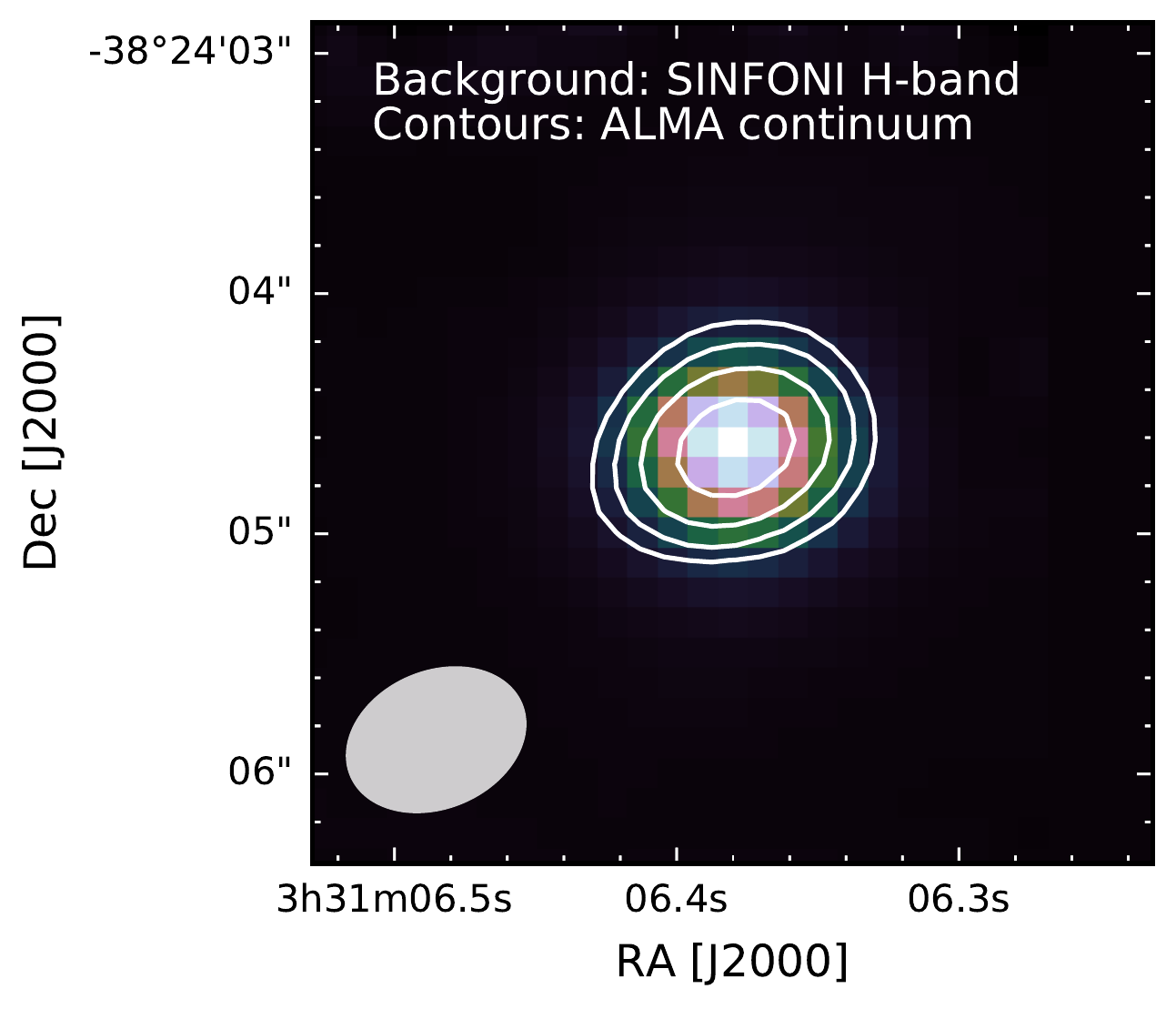}\\
\ \\
\ \\ 
 {\LARGE  (b) }
  \includegraphics[width=0.8\columnwidth]{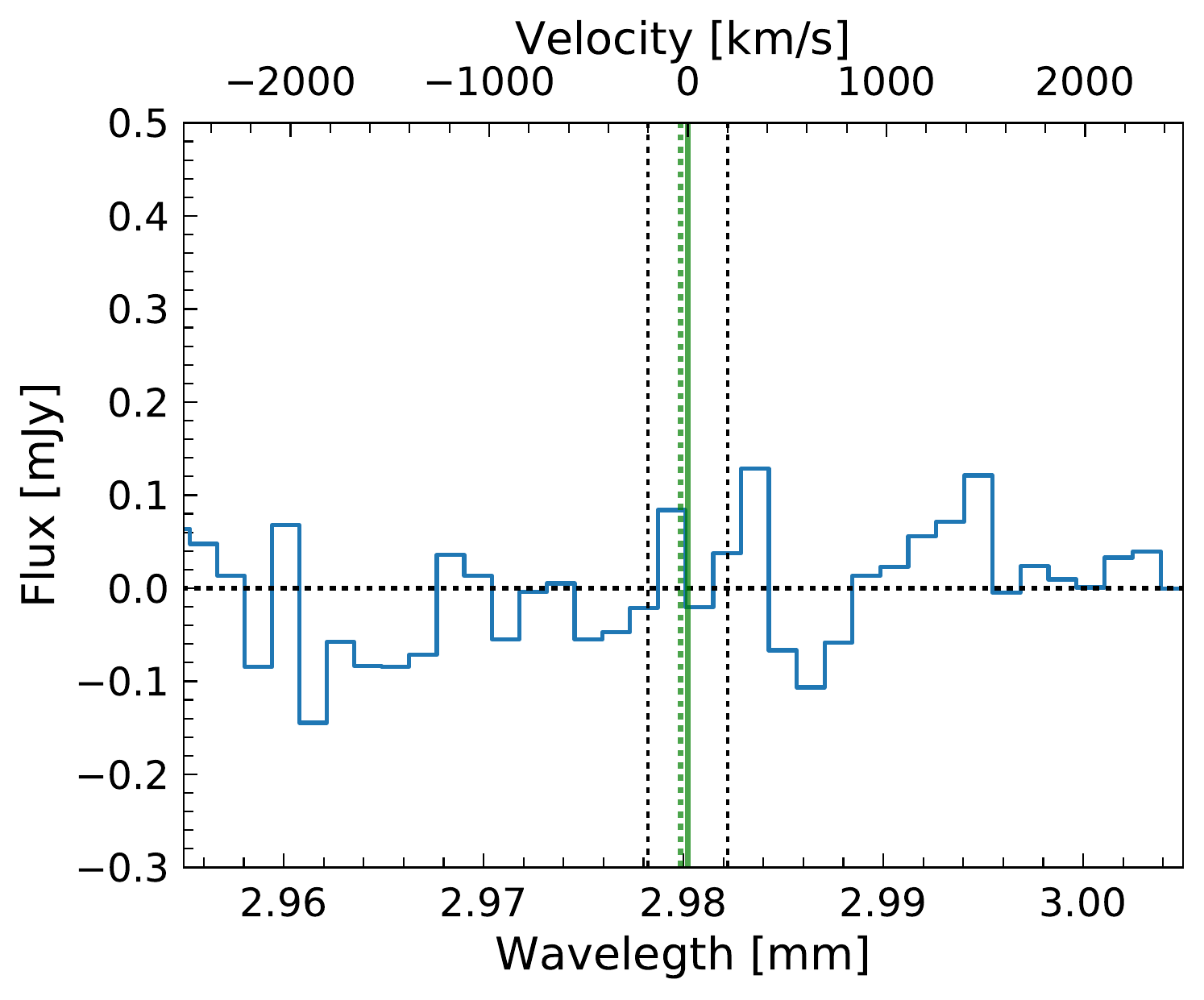}
  \quad \quad  {\LARGE  (c) }
  \includegraphics[width=0.8\columnwidth]{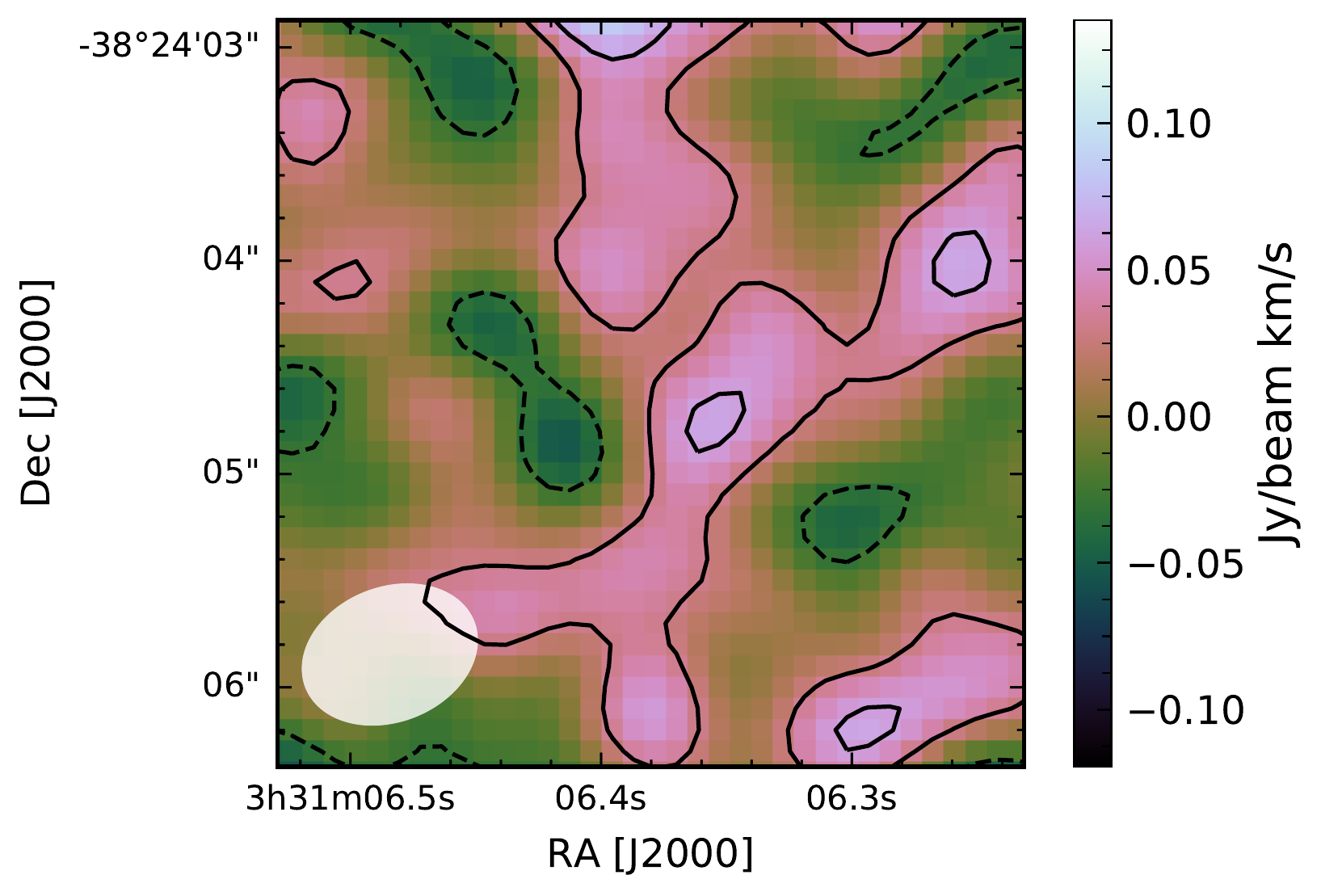}
 \caption{ \hbs:  (a) White contours show the continuum emission from \hbs\ at 3 mm with a beam of 0.62\arcsec$\times$0.46\arcsec (the synthesised beam is shown in the bottom-left corner). Contours are at the level of 25, 50,100 and 200 times the noise per beam (18 \ujy). The colour background image shows the continuum emission in H-band from SINFONI data of HB8903.(b) ALMA spectrum extracted from an aperture as large as the beam size and rebinned to 120 km/s. The vertical dashed and solid green line mark the expected positions for CO(3-2) emission line at the redshifted of the narrow \ha\ and \oiii\ component, respectively \citep{Carniani:2016}: the line is not detected.
 (c) CO(3-2) map obtained by integrating the cube under the two vertical dotted line indicated in the panel (b) (i.e. -200 km/s $<v<$ 200 km/). Negative and positive contours are in steps of $1\sigma$, which is  0.027 Jy/beam km/s.  The synthesised beam is shown in the bottom-left corner.} 
\label{fig:hb8903}
\end{figure*}


Figure~\ref{fig:hb8903}a shows the 3~mm continuum emission of HB8903, which is detected with  a high S/N of approximately $300$. 
The total 3~mm flux density of the QSO is $5.738\pm0.018$ mJy (including calibration uncertainties) that is, about 30 times higher than that measured in the  two previous QSOs (see Table~\ref{tab:alma}). 
 From a 2D-Gaussian fitting in the UV-plane we estimate a beam-deconvolved size of about 0.1\arcsec.

HB8903 has been identified as  a radio-loud QSO \citep{Shemmer:2004}  with a luminosity  of ${\rm log_{10}}(L_{8.4}/{\rm W Hz^{-1}}) \simeq 27.7$ at 8.4 GHz \citep{Healey:2007}. Therefore, the 3~mm flux is probably dominated by synchrotron emission which does not allow us an estimate of the far-infrared emission associated with the dust. 
Similarly to the analysis performed in the previous sections, we estimate an upper limit on the dust mass of $M_{\rm dust}$ = 2-3$\times10^{10}$ \msun, depending on dust temperature. 

The CO(3-2) line is not detected at the location of the QSO as shown in the  panels (b) and (c) of Fig.~\ref{fig:hb8903}. 
To estimate a $3\sigma$ upper limit on the line flux, we assume a line width as large as that observed in \lb\ (FWHM = 400 km/s) yielding $S_{\rm CO(3-2)}\Delta v < 0.08$ Jy km/s (Table~\ref{tab:alma}). 
We then infer an upper limit on the molecular gas mass of M$_{\rm gas}$(\aco=0.8) $< 0.2\times10^{10}$~\msun\ and M$_{\rm gas}$(\aco=4) $< 1.0\times10^{10}$~\msun\ for the two different CO-to-H$_2$ conversion factor, respectively.  

In the ALMA field of view of HB8903, we  also detect two additional sources with likely molecular line emission: one located 4.9\arcsec ($\sim$40 kpc at z = 2.44) to the north-east of the QSO and the other 16.6\arcsec ($\sim$140 kpc at z = 2.44) to the south-west of the QSO. The properties of these galaxies are discussed in Appendix~\ref{sec:appA}. 
As already observed in LBQS0109, even the QSO HB8903 could be located in a overdensity at $z\simeq2.4$ (see Sect.~\ref{sec:overdensity})

\section{Lack of molecular gas}
\label{sec:lack}

\begin{figure}
 \includegraphics[width=\columnwidth]{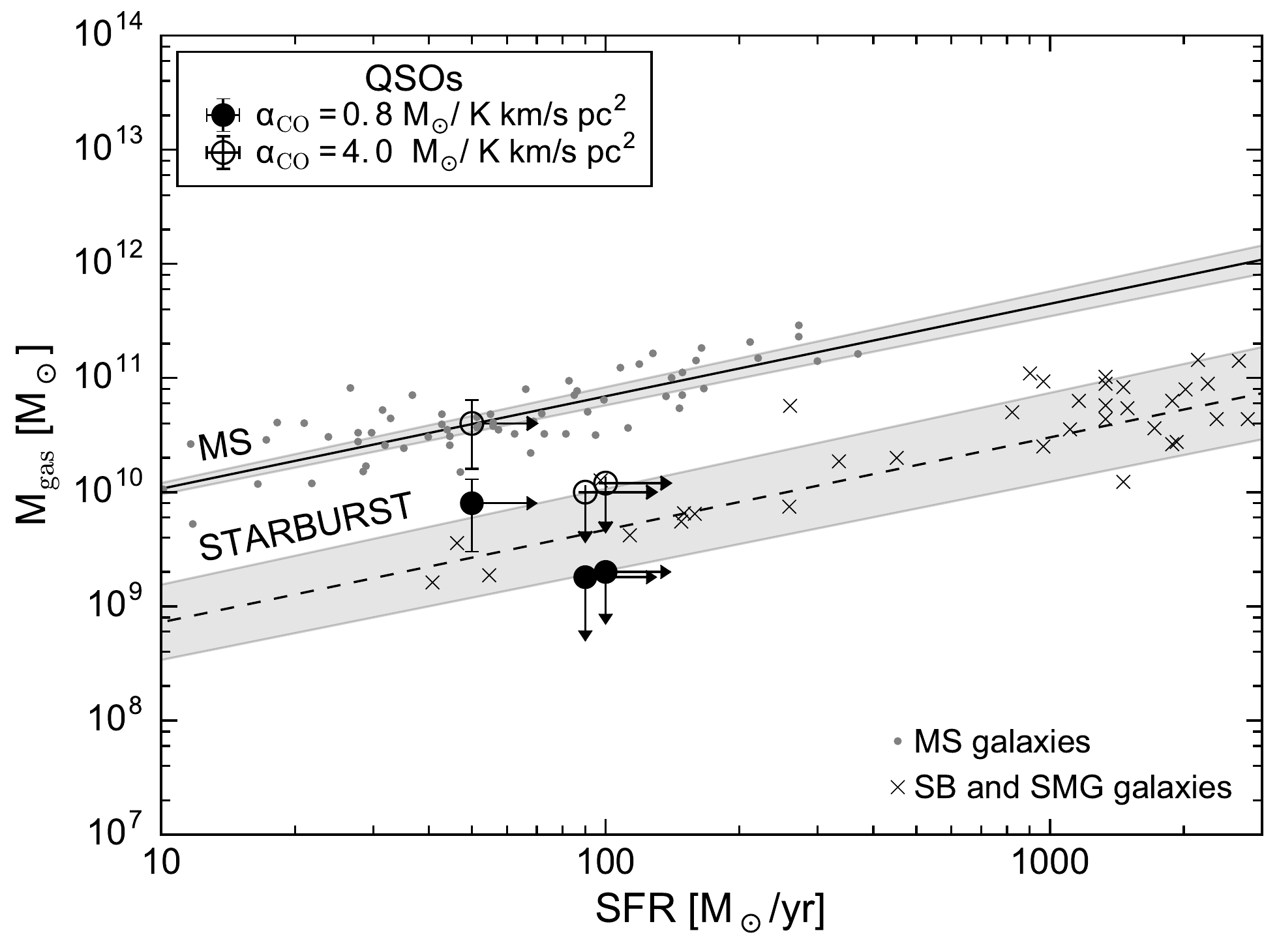}
 \caption{ Inverse, integrated version of the Kennicutt-Schmidt relation between SFR and molecular mass gas. The solid black line is the best-fit relation for MS galaxies and the dashed shows the relation of the starburst galaxies \citep{Sargent:2014}. 
 Filled black circles corresponds to our three QSOs assuming a conversion factor \aco=4 and open circles are derived supposing a \aco=0.8}
 \label{fig:mstar_sfr}
\end{figure}

Figure~\ref{fig:mstar_sfr} shows the best-fit relations by \cite{Sargent:2014} between molecular gas mass and SFR  for massive (\mstar $> 10^{10}$ \msun) main-sequence galaxies and starburst at low and high redshift ($z$<4). 
The SFRs of  our three QSOs  are estimated from the narrow  \ha\ emission \citep{Cano-Diaz:2012, Carniani:2016} assuming a Chabrier initial mass function \citep{Chabrier:2003}. 
 Since the \ha\ emissions are not corrected for reddening, the inferred SFRs are lower limits for both targets. 

In the \mgas-SFR plane,  all three sources are  placed   below the relation extrapolated for star-forming galaxies. In fact, a main-sequence star-forming galaxy with SFR=50 \sfr\ has a molecular gas mass of \mgas=$4\times10^{10}$ \msun\ that is similar to \mgas(\aco=4) of \lb, but it is at least five times higher if we assume a conversion factor \aco=0.8 (see Table~\ref{tab:alma}).
The inferred molecular gas masses are comparable with the expectation based on \cite{Sargent:2014}.

We can estimate the depletion timescale, which is defined as the rate at which the gas is converted into stars: \tscale = \mgas/SFR. Because the SFR from \ha\ are lower limits, we infer an upper limit of \tscale<160-800 Myr  for LBQS0109,  
\tscale<30-120 Myr for 2QZJ0028 and \tscale<20-110 Myr for HB8903, depending on \aco. 
The depletion timescales are similar to those observed in starburst  and SMG galaxies \citep{Yan:2010, Bothwell:2013} and reddened QSOs  \citep{Banerji:2016} at $z\sim2.5$, suggesting that the three host galaxies may still be in a starburst phase and  the star formation activity is not affected by AGN-driven outflows.

However the low molecular gas mass may also indicate that a fraction of the gas reservoir 
is expelled away from the galaxy by AGN-driven feedback, and, at the same time, the fast winds induce 
high pressure in the rest of the gas, triggering star formation in the region unaffected by AGN activity \citep{Silk:2013}. This scenario is similar to that observed in the QSO XID2028 at redshift $z\sim1.5$, where the presence of ionised outflow has been observed through \oiii\ emission \citep{Brusa:2015, Perna:2015, Cresci:2015} and the  small gas reservoir, respect to a MS star-forming galaxy with similar \mstar, is explained by negative-feedback \citep{Brusa:2015a}. Also, \cite{Kakkad:2017} have recently reported a lower gas fraction for a sample of AGN at $z\sim1.5$ compared with a sample of galaxies without an AGN matched in redshift, stellar mass, and star-formation rate.
In addition, \cite{Fiore:2017} have found that  the molecular gas depletion timescale
and the molecular gas fraction of a sample of 15 galaxies hosting powerful AGN driven winds are between three and ten times smaller than
those of main-sequence galaxies with similar star-formation rate, stellar mass and redshift.
According to such negative-feedback scenario, the molecular gas should be removed in the host region with the high velocity outflow.
Instead, the anti-correlation between CO emission and ionised outflow direction indicates  that a fraction of the gas has been already expelled from the galaxy.
Deeper ALMA observations will confirm this scenario also in \qz.

 Although the three QSOs have similar properties, such as SFR, \lagn, \mbh, the CO(3-2) at the systemic velocity of  the host galaxies is visible only in one of the three targets.
This discrepancy may be due to a different CO(3-2) excitation in the other two host galaxies. In \qz\ and HB8903 the CO(3~-~2) may be less excited than that in \lb\ and the current sensitivity is not sufficient to detect the emission line in the host galaxy.
In addition the  dense molecular clouds invested by AGN-driven wind develop Kelvin-Helmholtz instabilities \citep{Hopkins:2010, Ferrara:2016}. These instabilities  develop shocks responsible for higher gas excitation likely resulting into the strong CO(3-2) blueshifted emission that we observe in \qz.
A similar result is recently reported by \cite{Dasyra:2016} and \cite{Morganti:2015} who observed CO(4-3) and CO(2-1) emission in a local Seyfert galaxy, IC 5063.
Most of the CO(4-3) emission has been detected in the outflow regions, while the CO(2-1) is mainly emitted in the host galaxy location. In the outflow regions the CO(4-3)/CO(2-1) flux ratio approaches 16 ($\times3$ higher than that observed in the rest of the host galaxy ).  In this regard, we also note that CO observations in the distant Universe ($z>1$) show that the CO spectral line energy distribution of normal star-forming galaxies are less excited than those of  SMGs and QSOs. The ratio between mid-J and low-J CO transition measured in SMGs and QSOs is higher than that observed in normal galaxies by a factor $>1.5$ \citep{Carilli:2013, Gallerani:2014, Mashian:2015, Daddi:2015}.

\section{Overdensity}
\label{sec:overdensity}

ALMA observations have revealed the  presence of six companion sources within a projected distance $\sim 160$ kpc from the quasars (Appendix A).  In five out of the six sources we also detect a line emission  that may be identified with CO(3-2) transition at similar redshifts of the QSOs. 
The CO(3-2) lines in these  sources have luminosities of 0.6-23 K km/s pc$^2$, resulting in a molecular masses of $0.5-90\times10^{10}$~\msun\ that are even higher than those measured in the QSOs themselves. The molecular mass and the high SFR ($\sim1000$  \sfr), inferred from the continuum emission is comparable to those observed in starbursts and submillimetre galaxies (SMGs). In one case, the line emission is even spatially resolved by the ALMA beam and the gradient of velocity indicates a dynamical mass of $M_{\rm dyn}=2\times10^{11}\sin^2(i)$, which is similar to those observed in high-z SMGs \citep[e.g.][]{Carniani:2013}. 

Recent galaxy evolution models predict that the rate of galaxy mergers and interactions increases in the redshift range $1<z<3$, driving extreme starburst events and rapid accretion onto the massive black holes in the galaxy centre \citep{Di-Matteo:2005, Sijacki:2011, Valiante:2011}.
These predictions have been supported by new extragalactic surveys at millimetre and submillimetre ranges having uncovered a population of dusty star-forming galaxies at high redshift.
\cite{Silva:2015} found an overdensity of submillimetre galaxies in 17 out of 49 QSOs at redshift $z\sim2$. 
A similar scenario has been recently reported by \cite{Banerji:2016} who detected two millimetre-bright galaxies  within 200 kpc from a QSO at $z=2.5$. 
An overdensity system is also observed in BR 1202-0725, which is mainly composed by a QSO and a submillimetre galaxy  at $z\sim4.7$  \citep{Salome:2012, Wagg:2012,Carilli:2013b, Carniani:2013,Williams:2014}.
Overall, these results support the hypothesis that submillimetre galaxies and QSOs represent different stages of galaxy evolution after a merger \citep{Carniani:2013}.
The detections of massive companions sources in the ALMA field of view have been also observed in QSOs at higher redshifts \citep[$z\sim4.8-6$;][]{Trakhtenbrot:2017, Decarli:2017} indicating that major mergers are important drivers for rapid early SMBH growth.

The detections of these seredentipitous sources in our ALMA observations suggest that the three QSOs are located in a overdensity. Assuming that a SMBH of 10$^{10}$ \msun\ is associated to a dark halo of mass of  10$^{13}$ \msun\ \citep[e.g.][]{Ferrarese:2002}, we estimate a virial radius of about 500 kpc for the three QSOs. This is larger than the projected distance between the QSOs and the serendipitous companions. We therefore conclude that the QSOs  and the serendipitous sources may  represent a complex merging system at redshift $z\sim2.3-2.5$. Future  millimetre observations at different wavelength bands will confirm the redshift of the serendipitous galaxies  and their nature.

\section{Summary}
\label{sec:summary}

We have presented new ALMA 3mm observations aimed at mapping CO(3-2) in three $z\sim2.4$ quasars, \lb, \qz, and HB8903, showing evidence for ionised outflows quenching star formation \citep{Cano-Diaz:2012, Carniani:2015a, Carniani:2016}.
Below, we summarise the main results of this work: 

\begin{itemize}

\item The ALMA observations reveal the presence of serendipitous galaxies, three of those are detected both in  continuum (at 3 mm) and in  line emission, within a projected distance of 160 kpc from the QSOs. 
Assuming the emission line detected in these galaxies can identified with the CO(3-2) transition, we conclude that \lb\ and HB8903  reside in  overdense systems, as often found for QSOs at similar and higher redshifts.

\item The CO(3-2) emission at the systemic velocity of the QSO is detected in only one  of the three targets, that is, \lb. 
The CO profile has a velocity and line width consistent with the narrow \oiii\ and \ha\ components tracing SF in the host galaxy. 
In addition the CO emission is spatially resolved by the ALMA beam and is not symmetrically distributed around the location of the QSOs, but  absent or faint in the outflow region.
This is suggestive of a scenario in which the AGN-driven outflow is removing the ionised and molecular gas from the host galaxy.

\item In \qz\ we tentatively detect a faint CO(3-2) emission blueshifted by 2000 km/s relative to the redshift of the host galaxy and spatially coincident with the ionised outflow emission. If confirmed by follow-up observations, this CO emission may be tracing a  molecular cloud at high velocity that has been ejected away from the galaxy by AGN-driven outflows.
Also, our analysis would suggest that the molecular gas in the outflow region is more highly excited than the rest of molecular gas in the host galaxy.
An alternative interpretation to the outflow scenario could be that the detected CO emission is associated to a faint companion galaxy.
Future deeper ALMA observations of CO(3-2) and higher (or lower) rotation transition will be fundamental in confirming the redshift of this detection and analyse the excitation state of the molecular gas. If the outflow scenario will be confirmed, the new ALMA observations will allow us to estimate the molecular outflow mass rate and compare this value with that estimated from ionised gas in the same QSO. 

\item Assuming a \aco=0.8 \msun/ km/s pc$^2$, the inferred molecular gas mass in both host galaxies is clearly below what observed in  MS galaxies with similar SFR and consistent with those observed in other high-$z$ QSO and SMGs.

\end{itemize}

We conclude that AGN-driven outflows in our sample are removing ionised and molecular gas from the host galaxy and quenching the star formation. 
The interaction of the fast winds with the \lb\ host galaxy is clearly  visible  in the outflow region, where the CO emission is faint. 
 This result supports our previous studies  of these QSOs showing \ha\ emission quenched in the outflow region.

\section*{Acknowledgements}

This paper makes use of the following ALMA data: ADS/JAO.ALMA\#2013.0.00965.S; which can be retrieved from the ALMA data archive: https://almascience.eso.org/ alma-data/archive. ALMA is a partnership of ESO (representing its member states), NSF (USA) and NINS (Japan), together with NRC (Canada) and NSC and ASIAA (Taiwan), in cooperation with the Republic of Chile. The Joint ALMA Observatory is operated by ESO, AUI/NRAO and NAOJ.
SC and RM acknowledge financial support from the Science and Technology Facilities Council (STFC).
RM acknowledges ERC Advanced Grant 695671 ``QUENCH''.
MB acknowledges support from the FP7 Career Integration Grant ``eEASy'': Supermassive black holes through cosmic time: from current surveys to eROSITA-Euclid Synergies" (CIG 321913). 
CC acknowledges funding from the European Union's Horizon 2020 research and innovation programme under the Marie Sk\l{}odowska-Curie grant agreement No 664931. RS acknowledges support from the European Research Council under the European Union (FP/2007-2013)/ERC Grant  Agreement n. 306476. CF acknowledges funding from the European Union's Horizon 2020 research and innovation programme under the Marie Sk\l{}odowska-Curie  grant agreement No 664931.

\bibliographystyle{aa} 
\bibliography{bibliography_ALMA} 




\appendix

\section{Overdense systems}
\label{sec:appA}

\begin{figure*}
 \includegraphics[width=2.\columnwidth]{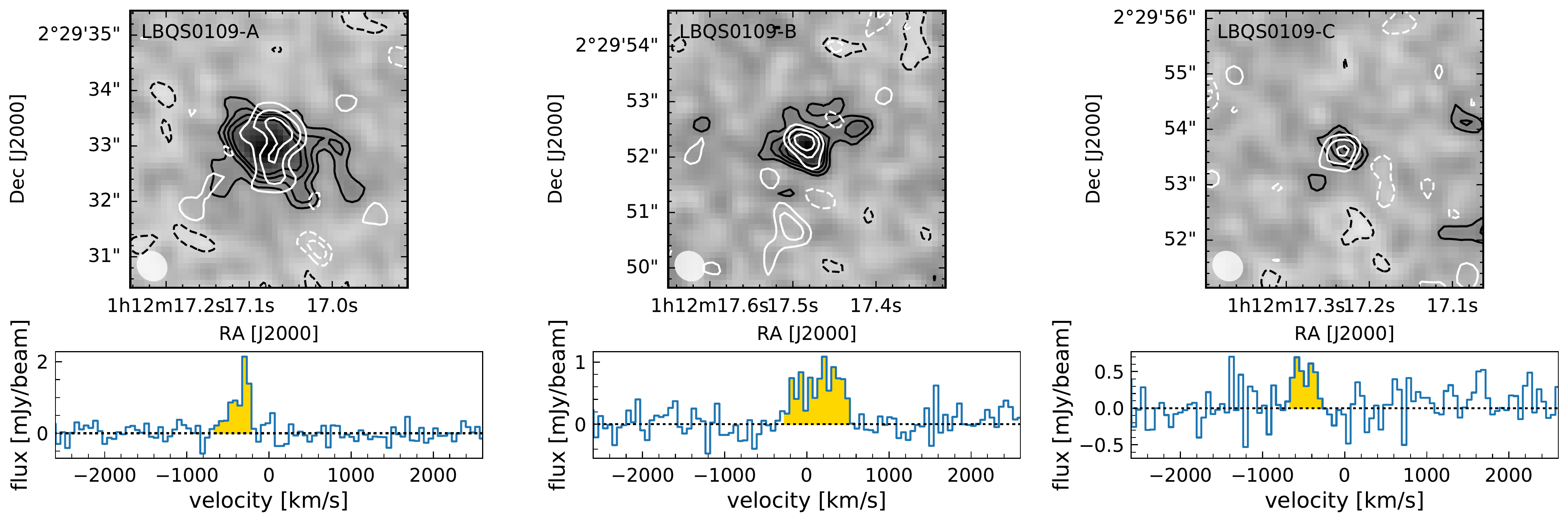}
 \caption{ CO(3-2) maps (top panels) of serendipitous sources detected in the \lb\ field of view. The maps are obtained collapsing under golden shade region shown in their respective spectra (bottom panels). The black contours are at levels of -2, 2, 3, 4, and 5  times the noise per beam in the same map (i.e. 0.06, 0.05, and 0.05 Jy/beam km/s, respectively for the three sources). The ALMA beam is shown in the bottom-right corner of each map.
 The white contours indicate the continuum emission at levels of 2(-2)$\sigma$, 3(-3)$\sigma$, and 4$\sigma$. The spectra in the bottom panels are extracted from the centre of the sources.
 The 0 km/s corresponds into the redshift of \lb.}  
 \label{fig:lbqs0109-sources}
\end{figure*}

\begin{figure*}
\centering
 \includegraphics[width=1.5\columnwidth]{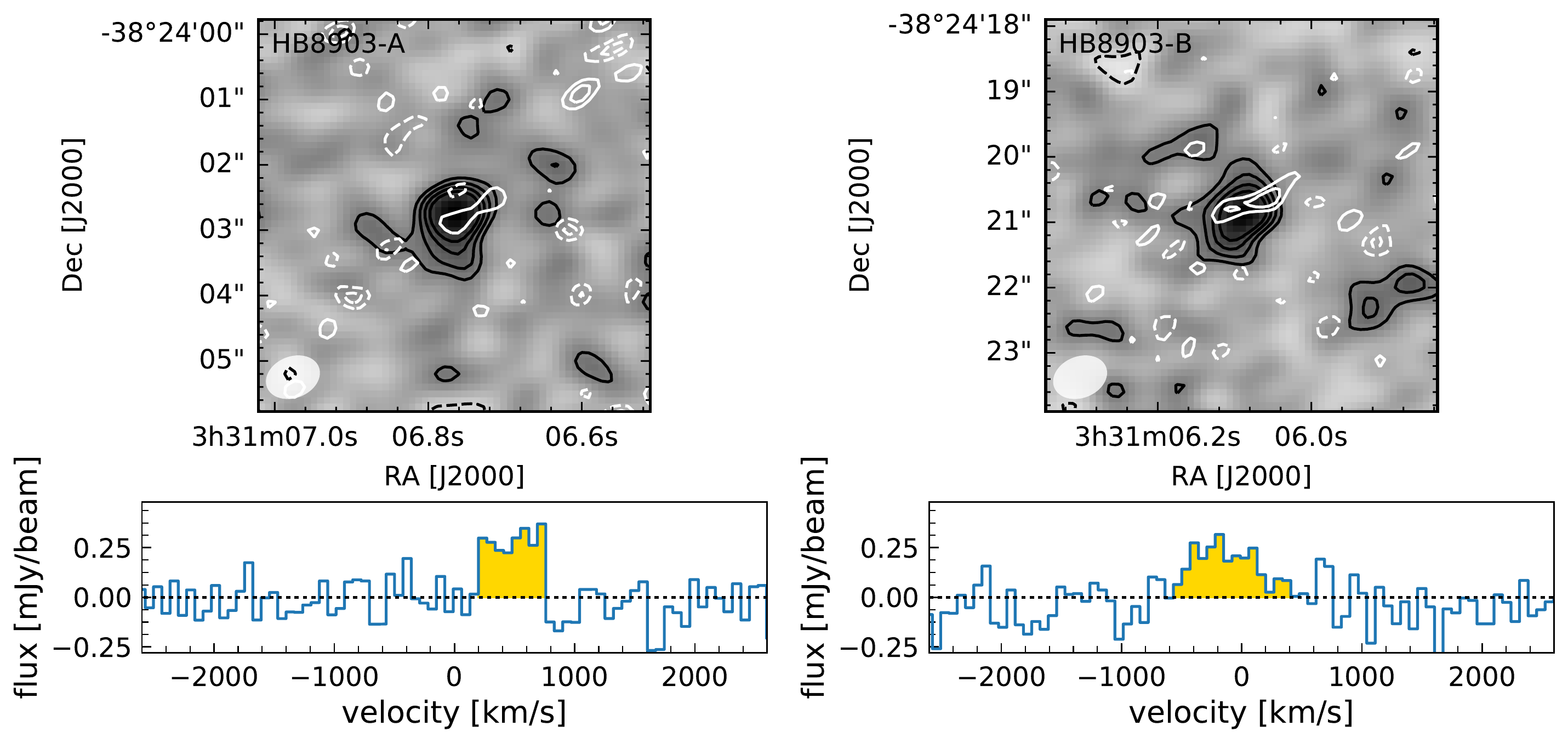}
 \caption{ CO(3-2) maps (top panels) of serendipitous sources detected in the HB8903 field of view. The maps are obtained collapsing under golden shade region shown in their respective spectra (bottom panels).  Black contours contours are  in steps of 1$\sigma$ = 0.03 Jy/beam km/s for the source A and 1$\sigma$ = 0.04 Jy/beam km/s for the source B, starting at $\pm2\sigma$ . The ALMA beam is shown in the bottom-right corner of each map.
 The white solid(dashed) contours indicate the continuum emission at levels of 2(-2) and 3(-3) times the noise per beam (18 \ujy/beam). The spectra in the bottom panels are extracted from the centre of the sources.
 The 0 km/s corresponds into the redshift of HB8903.  }
 \label{fig:hb8903-sources}
\end{figure*}

\begin{figure}
 \includegraphics[width=\columnwidth]{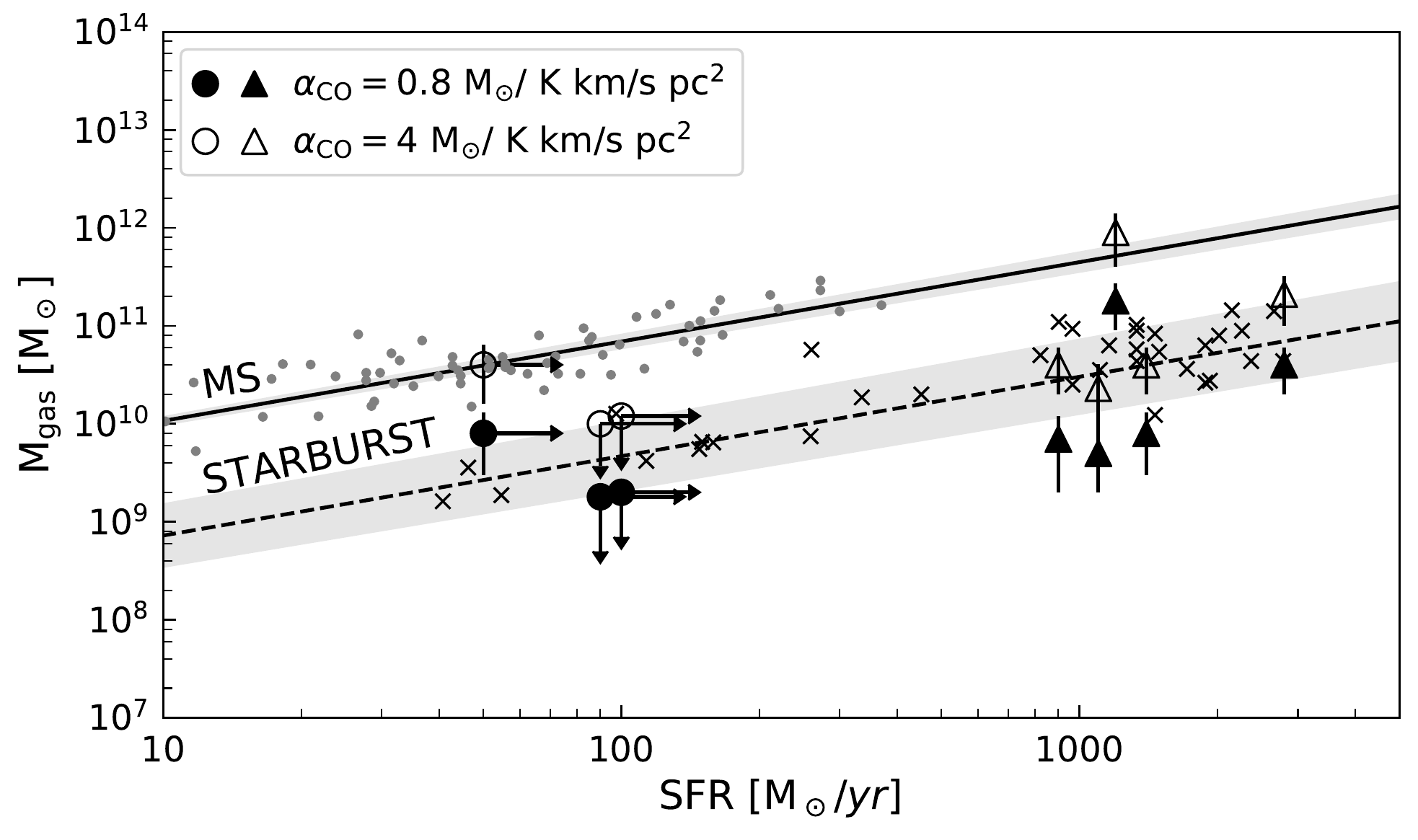}
 \caption{ Inverse, integrated version of the Kennicutt-Schmidt relation between SFR and molecular mass gas. The solid black line is the best-fit relation for MS galaxies and the dashed shows the relation of the starburst galaxies \citep{Sargent:2014}. 
 Circles and triangles correspond to the three QSOs and serendipitous sources, respectively. } 
 \label{fig:mgas_sfr_app}
\end{figure}

The ALMA observations serendipitously reveal line and continuum sources in the field of \lb\ and HB8903 quasars, (Fig.~\ref{fig:lbqs0109-sources} and \ref{fig:hb8903-sources}),  located within a projected radius of $\sim~160$~kpc from the centre of the two QSO. From the continuum emission we infer a SFR$>$900 \sfr\ for all sources and a M$_{\rm dust}~=~1-6 \times10^8$ \msun.
In Table~\ref{tab:serendipity} we list the properties of these detections and, hereafter, we use the terms LBQS0109-A, -B ,-C, and HB8903-A, -B  to  refer to the serendipitous sources around LBQS0109 and  HB8903, respectively.
If the  emission line are identified with CO(3-2) transition, the serendipitous sources are in a redshift range of $\Delta z = 0.005$ ($|\Delta v|<500$ km/s) relative to the respective QSOs. 
We estimate a CO(3-2) luminosity of 0.6-23$\times10^{10}$ K km/s pc$^2$ for these sources and a molecular gas mass of 0.5-90$\times10^{10}$ \msun, depending of \aco\ (Table~\ref{tab:serendipity}), which is higher than molecular gas masses inferred for the three QSOs. The SFR and M$_{\rm gas}$ of these serendipitous sources are consistent with those observed in 
SMGs (Fig.~\ref{fig:mgas_sfr_app}). 
Assuming a dark halo mass of $10^{13}$ \msun, which is reasonable for a galaxy hosting a massive BH of mass $10^{10}$ \msun, we note that the serendipitous  galaxies are  within the virial radius ($\sim500$ kpc) of the central QSO. Both \lb\ and HB8903 could represent an overdense system at $z\simeq2.4$.


\begin{figure}
 \includegraphics[width=\columnwidth]{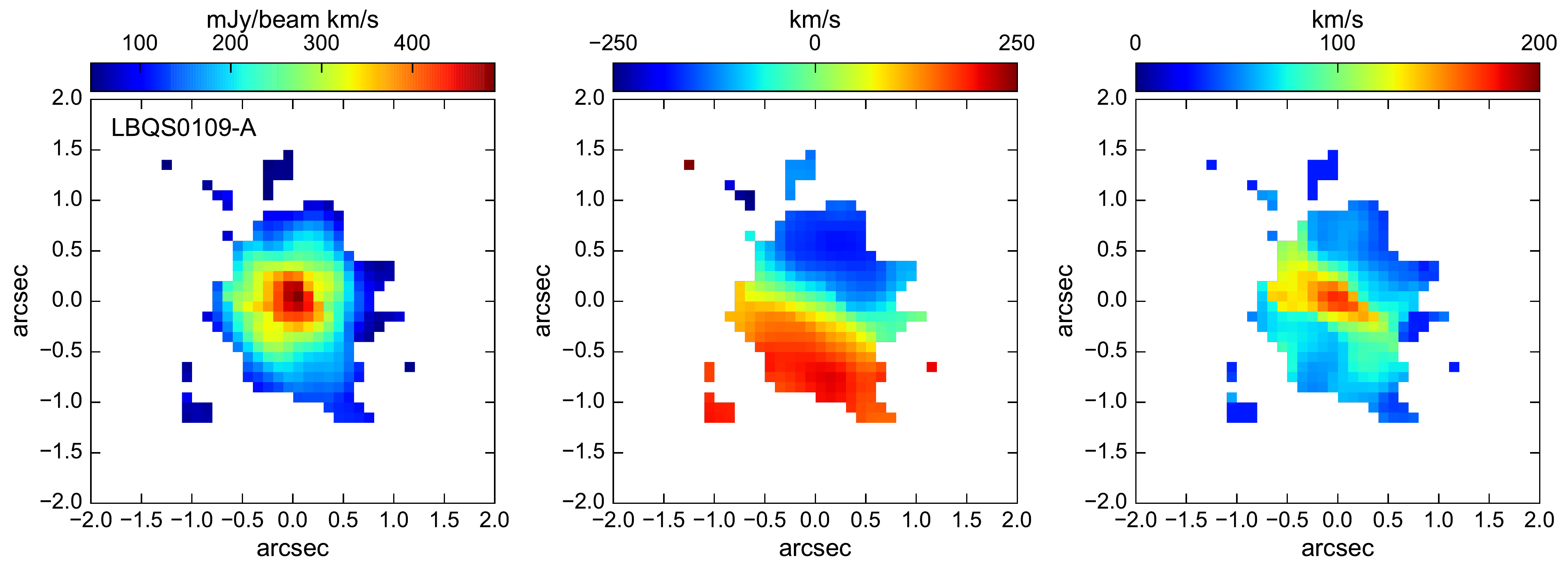}
 \caption{ CO(3-2) flux, median velocity, and velocity dispersion maps of LBQS0109-A. The maps are obtained by selecting pixel with S/N>3. The velocity map shows a gradient of velocity towards the  north-south  direction. }
 \label{fig:galaxyA_kinematic}
\end{figure}

From a 2D Gaussian fitting we estimate the size of the CO(3-2) emission in all serendipitous galaxies and  two  out of five sources turn out to be spatially resolved (LBQS0109-A and LBQS0109-B).
However, because the S/N-per-pixel in the LBQS0109-B is too low ($<10$), we perform  a pixel-by-pixel kinematic analysis  only on LBQS0109-A. 
The results of the kinematic analysis are reported in Fig.~\ref{fig:galaxyA_kinematic} that shows a clear north-south  velocity gradient of $\sim500$ km/s over $\sim1\arcsec$ (8.3 kpc).
If it were due to simple rotation, this would imply a dynamical mass without inclination angle $i$ correction of $2\times10^{11}\sin^2(i)$ \msun.

\begin{figure}
\centering
 \includegraphics[width=0.8\columnwidth]{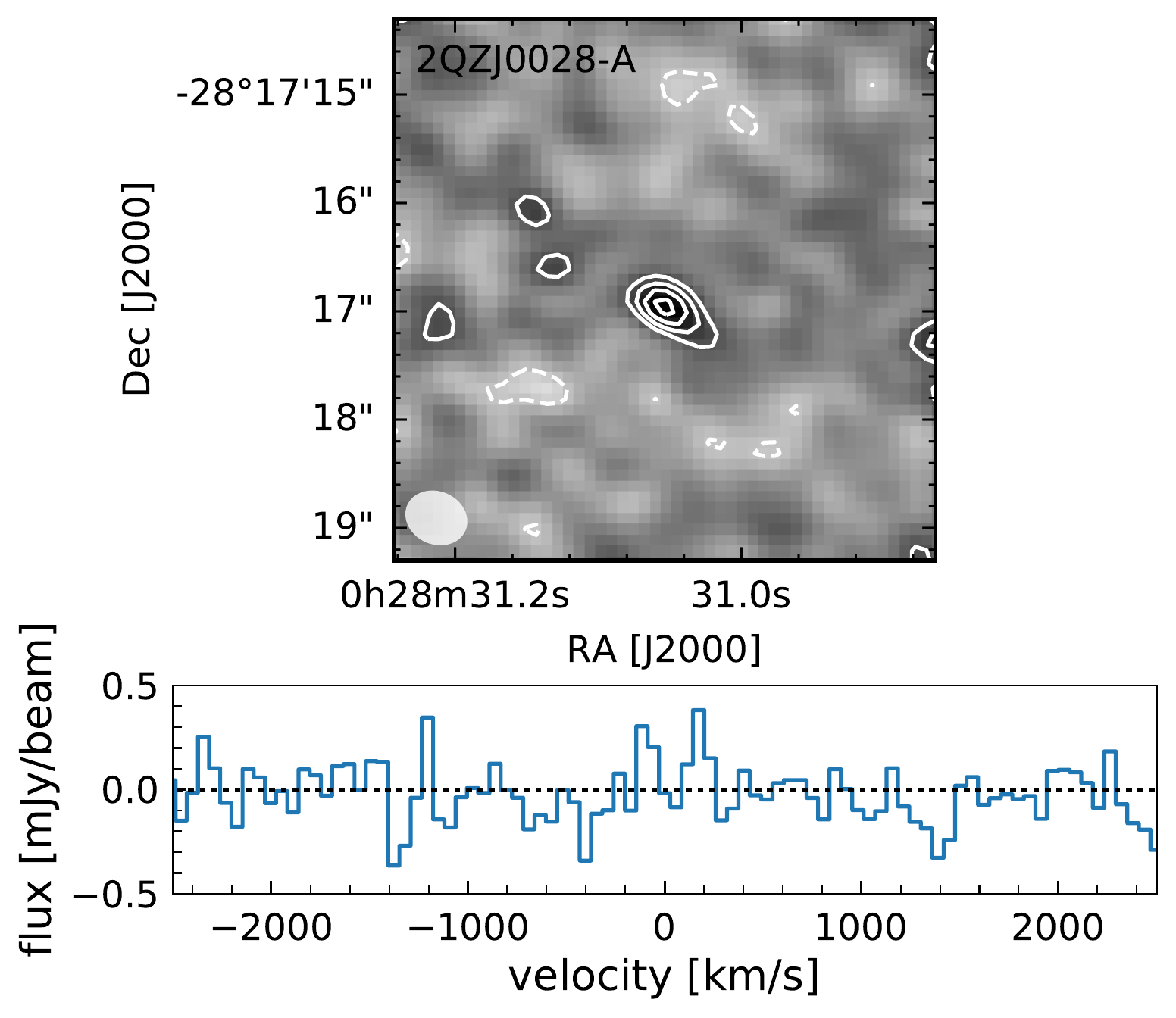}
 \caption{ Continuum map at 3~mm of the  serendipitous source detected in the field of \qz. Contours are at of -2, 2, 3, 4, 5 times of the continuum rms (12 \ujy/beam). The ALMA beam is shown in the bottom-right corner of each map.}
 \label{fig:2qz_source}
\end{figure}

We do not detect any significant (S/N$>5$) line emission in the ALMA field-of-view of \qz\ except the line emission candidate spatially offset by 0.2\arcsec\ discussed in Sect.~\ref{sec:possible_detection}. 
We observe a continuum emission with a S/N=5 located at $\sim14\arcsec$ from the QSO. 
The continuum emission (Fig.~\ref{fig:2qz_source}) is not spatially resolved and its flux density at 3~mm is $61\pm12$ \ujy. 

The comparison of our ALMA detections with  number counts studies at similar wavelengths indicates that all QSOs reside in a significant over-density. 
In fact, theoretical results obtained by simulations and semi-analytical models predict a number counts at 3 mm of N(S$>100$\ujy)$\simeq100\deg^{-2}$ \citep{Takeuchi:2001, Cai:2013}, yielding to a N(S$>100$\ujy)$\simeq0.02 \ {\rm FOV^{-1}_{\rm ALMA}}$ where ${\rm FOV_{\rm ALMA}}$ is the field-of-view area of our ALMA maps ($\sim2500$ arcsec$^2$).
Similar number counts values are also observed in recent ALMA studies at 1.3 mm \citep[e.g.][]{Carniani:2015}. 
The number of serendipitous sources detect in our ALMA maps is larger than that expected by simulation and millimetre surveys, hence supporting that the three QSOs are located in a overdense region.

\begin{sidewaystable}
\caption{Properties of serendipitous sources.}           
\label{tab:serendipity}      
\centering          
\begin{tabular}{l c c c c c c }    
\\
\hline
 &  LBQS0109-A & LBQS0109-B & LBQS0109-C & HB8903-A & HB8903-B & 2QZJ0028-A \\

\hline
\\
RA   &  01:12:17.08  &  01:12:17.49  & 01:12:17:23 & 3:31:06.77 & 3:31:06.09 & 00:28:31.812 \\
DEC  &  2:29:32.99 &  2:29:52.26  & 2:29:63.64 & -38:24:02.79 & -38:24:20.95 & - 28.16.51.616\\
S$_{\rm 3mm}$ [\ujy] & $115\pm12$  &  $52\pm12$ & $48\pm12$ & $36\pm18$ & $54\pm18$ & $61\pm12$ \\

M$_{\rm dust}$ [10$^{8}$ \msun]$^{a}$ & 4-6 & 1-3 & 1-3 & 1-2 & 2-3 & 2-3  \\

SFR [\sfr]$^{a}$ & 2800 & 1200 & 1100 & 900 & 1400 & 1500 \\
$\lambda_{\rm CO(3-2)}$ [mm] & $2.90642\pm0.00008$ & $2.9112\pm0.0003$ & $2.9045\pm0.0003$  & $2.9779\pm0.0006$ &  $2.9843\pm0.0005$ & - \\

FWHM$_{\rm CO(3-2)}$ [km/s] &  $190\pm20$ & $590\pm70$& $300\pm20$ & $400\pm100$ & $400\pm100$ & -\\

S$_{\rm CO(3-2)} \Delta v$ [Jy km/s] & $1.72\pm0.06$ & $7.16\pm0.05$ &  $0.19\pm0.05$ & $0.27\pm0.03$ &  $0.32\pm0.04$ & -\\

L$^\prime_{\rm CO(3-2)}$ [$10^{10}$ K km/s pc$^2$] & $5.2\pm0.2$ & $22.9\pm0.2$ & $0.58\pm0.15$ &  $0.88\pm0.09$ & $1.04\pm0.13$ & -\\

L$_{\rm CO(3-2)}$ [$10^{7}$ \lsun] & $6.9\pm0.2$ & $29.1\pm0.2$ & $0.8\pm0.2$ &  $1.26\pm0.13$ & $1.38\pm0.17$ & -\\


M$_{\rm CO}$(\aco=0.8) [$10^{10}$ \msun]$^{b}$ &  $4\pm2$ & $18\pm9$ &  $0.5\pm0.3$ & $0.7\pm0.4$ & $0.8\pm0.5$ & - \\

M$_{\rm CO}$(\aco=4) [$10^{10}$ \msun]$^{b}$ &  $21\pm11$ & $90\pm50$ &  $2.3\pm1.7$ & $3\pm2$ & $4\pm2$ & - \\
\\
\hline

\end{tabular}   
\tablefoot{ $^{a}$ Under the assumption that the continuum emission at 3~mm is completely associated  to thermal dust continuum emission.
We assume a ${T_{d}}$=40-60 K and a $\beta=2.0$. $^{b}$ 
Under the assumption that the line detection is associated to the CO(3-2) transition and assuming a \rco=$1.0\pm0.5$. The statistical errors associated to the molecular gas include \rco\ uncertainties.}
\end{sidewaystable}

\section{Reliability of the  CO(3-2) line detection in \qz.}
\label{sec:appB}

In Sect.~\ref{sec:possible_detection} we show the detection of a faint line emission spatially offset by 0.2\arcsec\ from the centroid of  continuum emission at 3~mm  and blusefhited of $\sim-2000$ km/s respect to the redshift of \qz. In this section we discuss more in detail the significance of this candidate line  emission.

\begin{figure}
\centering
 \includegraphics[width=0.8\columnwidth]{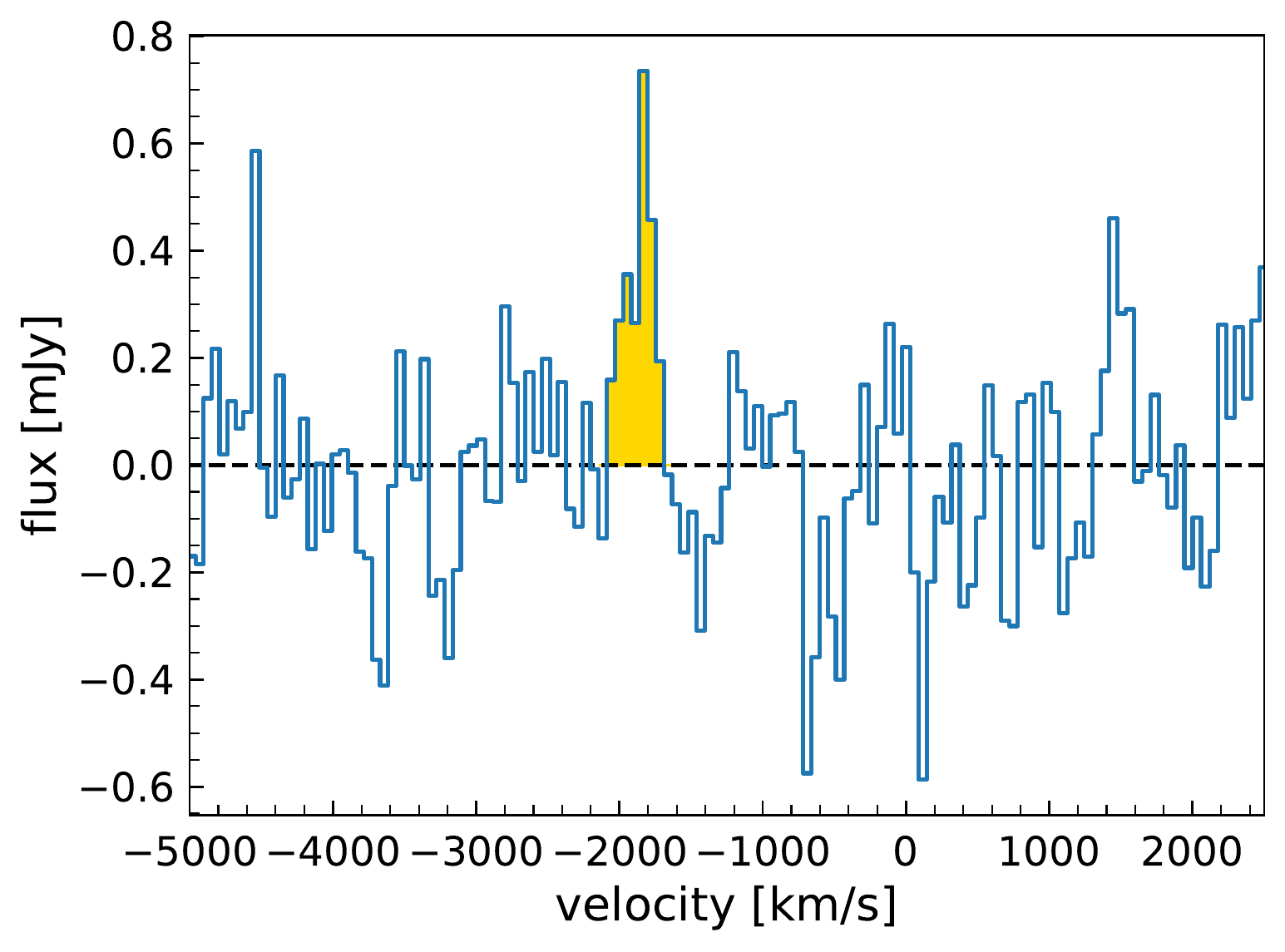}
 \caption{ ALMA spectrum extracted from the emission peak of the line detection, spatially offset by 0.2\arcsec\ relative to the location of \qz. The spectrum is rebinned to 60 km/s.}
 \label{fig:spectrum_outflow}
\end{figure}

We checked whether negative sources are detected with the same significance or not.
We performed a blind search for positive and negative line emitters within the ALMA primary beam area  and within a velocity range $|v|< 2000$ km/s relative to the redshift of QSO. We searched for line emitters with line width 
ranging from 200 km/s to 500 km/s. 
We extracted 16 positive and 10 negative emission line with a level of confidence higher 5$\sigma$. The number of positive peaks at S/N$>$ 5 is 50\% higher than the number of negative peaks, hence it indicates that few of these detections could be not noise fluctuations. In Fig.~\ref{fig:spectrum_outflow} is shown the spectrum of the blueshifted line extracted at the location of the emission peak. Focusing on the spectrum, we note that there are no negative peaks with a level of confidence above $3\sigma$ suggesting that the positive blueshifted line, which is detected with a S/N$>4$, is not due noise fluctuations. These tests and the coincidence that  both velocity and location of the line emitted are consistent with the ionised outflow support the reliability of this detection. 


\end{document}